\documentclass[aps,prx,reprint,floatfix,superscriptaddress,footinbib,twocolumn]{revtex4}
\usepackage{xcolor}
\usepackage{microtype}
\usepackage{mathrsfs}

\usepackage[bookmarks=true,
bookmarksnumbered=false, 
bookmarksopen=false, 
colorlinks=true,
linkcolor=webblue]{hyperref}
\definecolor{webblue}{rgb}{0, 0, 0.5} 
\hypersetup{colorlinks=true,urlcolor=webblue,citecolor=webblue}
\usepackage{graphicx}
\usepackage{subfigure}
\usepackage{url}
\usepackage{hyperref}
\usepackage{inconsolata}
\usepackage{amsmath}
\allowdisplaybreaks
\usepackage[capitalize]{cleveref}
\usepackage{braket}
\usepackage{outlines}
\usepackage{enumitem}
\usepackage{soul}
\usepackage{color}

\usepackage{bm} 

\usepackage[utf8]{inputenc}
\usepackage{siunitx,booktabs}
\usepackage{multirow}
\usepackage{amssymb}

\usepackage{scalerel}
\usepackage{relsize}

\usepackage{soul}
\usepackage{tabularx}

\usepackage{amsmath}
\usepackage{amssymb}
\usepackage{bbm}
\usepackage{braket}
\usepackage{xcolor}

\usepackage{comment}
\usepackage{pifont}

\begin{document}

\title{Group theory method for extracting order parameters\\ from scanning tunneling microscopy data}

\author{Julian Ingham}
\email[]{ji2322@columbia.edu}
\affiliation{Department of Physics, Columbia University, New York, NY, 10027, USA}

\author{Yu-Xiao Jiang}
\affiliation{Laboratory for Topological Quantum Matter and Advanced Spectroscopy (B7), Department of Physics, Princeton University, Princeton, New Jersey, USA}

\author{M. Zahid Hasan}
\affiliation{Laboratory for Topological Quantum Matter and Advanced Spectroscopy (B7), Department of Physics, Princeton University, Princeton, New Jersey, USA}

\author{Harley D.~Scammell}
\affiliation{School of Mathematical and Physical Sciences, University of Technology Sydney, Ultimo, NSW 2007, Australia}

\date{\today}

\begin{abstract}
Scanning tunneling microscopy (STM) is a powerful local probe of correlated electronic states. Here we present a group theoretical framework for the analysis of STM data, filtering STM images into components which provide a real space mapping of the local symmetry properties of the underlying density of states.  Using this formalism, we show that certain kinds of symmetry breaking are impossible to resolve in the first Brillouin zone, due to symmetry restrictions we term ``Bragg peak extinctions'' in analogy with related ideas in x-ray crystallography. We show extinct patterns of symmetry breaking can be resolved using sub-unit cell structure, and develop methodological details for the accurate extraction of this symmetry information.  We illustrate our results on synthetic STM data for $2\times 2$ charge density waves on the kagome lattice, and on topographic data for kagome metal ScV$_6$Sn$_6$. Our results provide a powerful method for extracting symmetry insights from STM data, and provide constraints on when and how certain ground states are experimentally observable.
\end{abstract}

\maketitle

\section{Introduction} 

A key question in the study of correlated electronic materials is why and how they deviate from a conventional symmetric metal. In particular, symmetry serves as a fundamental guiding principle in the characterisation of novel states of matter, placing strong constraints on the response of a material to external perturbations \cite{anderson2018basic}. For instance, the breaking of time-reversal symmetry (TRS) permits a non zero Hall conductivity, the breaking of inversion symmetry results in optical activity, and the breaking of continuous symmetries results in generalised rigidity and dissipationless transport as in superconductors.

A range of experimental probes exists to detect specific symmetry breaking patterns, such as translational symmetry breaking via diffraction techniques \cite{comin2016resonant, osborn2025diffuse, le2023tracking, fawcett1988spin, zaliznyak2014neutron}, or TRS breaking through magneto-optical effects \cite{kapitulnik2009polar}. Many of these techniques are constrained by symmetry-imposed selection rules -- for instance, Raman spectroscopy probes even-parity symmetry breaking excitations, but is blind to those of odd-parity \cite{devereaux2007inelastic}. As a consequence, it is challenging to simultaneously observe evidence the breaking of multiple symmetires in a single experiment. The need to simultaneously probe multiple symmetry breaking patterns is particularly pressing in correlated systems where multiple intertwined orders arise, alongside cascades of transitions that may involve combinations of mirror, rotational, and translational symmetry breaking \cite{fradkin2015intertwined}.

\begin{figure}[t]
\centering
\includegraphics[width = 0.9\columnwidth]{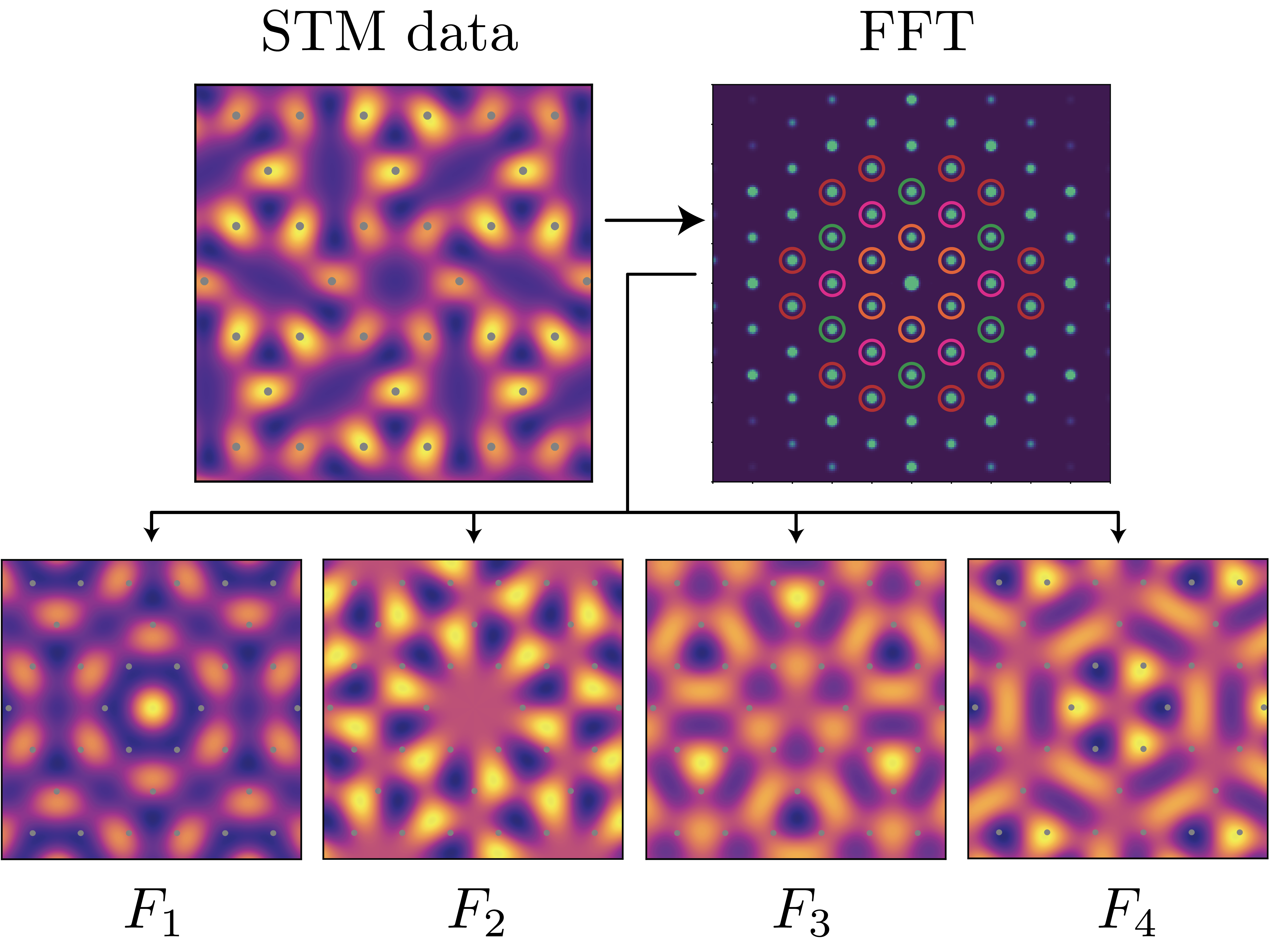}	
\vspace{-0.28cm}
\caption{\textbf{Group theory method for STM data.} Top left: synthetic STM data for $2\times 2$ CDW order on a kagome lattice. Top right: the FFT shows CDW Bragg peaks in the first (orange circles), second (magenta circles), and third (red circles) Brillouin zones, as well as the Bragg peaks of the lattice (green circles). Using group theory methods, the Bragg peaks can be projected into combinations with definite symmetry properties; inverse Fourier transforming filters the image into components corresponding to  possible order parameters $F_{1,2,3,4}$.}
\label{f:fig1}
\vspace{-0.65cm}
\end{figure}

Here we focus on scanning tunneling microscopy (STM), which permits a direct measurement of the local density of states (LDOS) at the surface of a material; understanding the symmetries present in STM data promises insight into the electronic ground state \cite{yazdani2016spectroscopic, nuckolls2024microscopic}. Previous studies have shown that symmetry-based techniques can be useful in analysing STM data; a fascinating example is the correlated insulator at half-filling in moir\'e graphene, a $\sqrt{3}\times \sqrt{3}$ Kekul\'e charge density wave (CDW) \cite{Nuckolls2023}. The Kekul\'e CDW was analysed via an innovative approach of decomposing the Bragg peaks of the STM data into symmetry channels in Fourier space, and then inverse Fourier transforming back to real space. The result is a set of images filtered into components with different symmetry properties, producing a real space map of the different symmetry-allowed components of the order parameter. Conceptually related ideas have been applied to study domains between different CDW configurations in NbSe$_2$ \cite{yoshizawa2024visualization},  image the CDW in TiSe$_2$ \cite{kim2024origin, wan2024ising}, and study intervalley scattering due impurities in graphene \cite{dutreix2019measuring}.

Yet despite these prior works, such a connection has not until now been made precise. Here, we establish and develop the connection between spatial symmetries and STM data, providing a general mathematical scheme for filtering the LDOS into components with distinct symmetry properties (Fig. \ref{f:fig1}), by generalising the ideas of Ref. \cite{Nuckolls2023} using methods from group theory. Surprisingly, we find selection rules which dictate that signatures of certain symmetry breaking patterns cannot be imaged in the first Brillouin zone; signatures of their rotational and mirror symmetry breaking only occur in higher Brillouin zones. We term these absent Bragg peaks ``Bragg peak extinctions'', in analogy with closely related ideas in x-ray crystallography \cite{bienenstock1962symmetry,rabson1991space}. We provide explicit formulae to determine which symmetries can be imaged, and which Fourier space data must be used to reconstruct the order parameter symmetry. Along the way, we develop methodological insights into how to analyse STM data to minimise artefacts and errors, allowing us to maximise the precision of this decomposition.

We consider explicitly the examples of $2\times 2$ and $\sqrt{3}\times\sqrt{3}$ CDWs in hexagonal systems, and present formulae in the Supplementary Material (SM) that generalise to arbitrary lattices with and without commensurate CDWs. As proof of principle, we apply our approach to synthetic STM data for translationally invariant and CDW orders on the kagome lattice -- confirming our predictions and illustrating the power of our method in determining the symmetry of the order parameter. We also analyse topographic data for kagome metal ScV$_6$Sn$_6$ to showcase our data analysis techniques and illustrate best practices.

Surprisingly, our work is the first to establish a precise connection between symmetry breaking orders, as captured in a mean-field Hamiltonian, and the LDOS. These results serve as a powerful tool and useful constraint in analysing STM experiments on novel symmetry breaking states of matter, and are generalisable to other techniques such as x-ray and neutron scattering.

\section{Results}

\subsection{Symmetry decomposition}
Starting from STM data $s(\mathbf r)$, which represents the LDOS measured at the surface of a material, we define the local Fourier component following the notation of Ref.~\cite{Nuckolls2023},
\begin{align}
\label{main:stm}
A(\mathbf Q_i, \mathbf r) 
=\int d^2 \mathbf{r}^{\prime} \, e^{-i \mathbf Q_i \cdot \mathbf r^{\prime}} s\left(\mathbf r^{\prime}+\mathbf r\right) w(\mathbf r')\,.
\end{align}
This expression captures the spatially resolved Fourier component of $s(\mathbf r)$ at wavevector $\mathbf Q_i$, centred at position $\mathbf r$, and weighted by a rotationally symmetric window function $w(\mathbf r')$. The window plays a critical role in suppressing spurious symmetry breaking, as we shall show. Different choices of $\mathbf Q_i$ are appropriate depending on the type of charge order being probed: taking $\mathbf Q_i = \mathbf M_i$ or $\mathbf Q_i = \mathbf K_i$ in hexagonal systems allows one to examine $2 \times 2$ or $\sqrt{3}\times \sqrt{3}$ CDWs respectively. Choosing $\mathbf Q_i = \mathbf G_i$, i.e. the Bragg peaks of the underlying lattice, allows one to analyse translationally symmetric orders.

The set of complex Fourier components $\{A(\mathbf Q_i, \mathbf r)\}$ can be decomposed into linear combinations transforming as so-called irreducible representations (irreps) of a material's space group -- i.e. the symmetries of the Hamiltonian under reflections, rotations, and translations (see SM Sec. \ref{supp-rep}). These irreps are the basic building blocks of symmetry analysis and play a central role in classifying phase transitions: in Ginzburg–Landau theory, an order parameter must transform as a single particular irrep of the parent symmetry group. In what follows, we show how to decompose $\{A(\mathbf Q_i, \mathbf r)\}$ into irreps, and how the appearance of a non-trivial irrep signals the breaking of the corresponding symmetry via an order parameter. This provides a direct link between STM data and an underlying order parameter.

To decompose these complex Fourier components into combinations distinguished by symmetry, we introduce concepts from group representation theory; the SM presents a review of the necessary mathematics. The group of all spatial transformations which leave a lattice invariant is the space group $\mathscr{S}$, consisting of all translations $\mathscr{T}$, and the point group $\mathscr{G}$. The point group can be viewed as the factor group of the space group, $\mathscr{G}=\mathscr{S} / \mathscr{T}$. The translation subgroup $\mathscr{T}$ is generated by $T\left(\mathbf{a}_{1}\right)$ and $T\left(\mathbf{a}_{2}\right)$, corresponding to the elementary lattice vectors $\mathbf{a}_{1}$ and $\mathbf{a}_{2}$, e.g. $\mathbf{a}_1=(1,0)$ and $\mathbf{a}_1=(-\tfrac{1}{2},\tfrac{\sqrt{3}}{2})$ for hexagonal lattices. Translationally invariant order parameters are classified as irreducible representations of the point group; to analyse states with finite wavevector $\mathbf{Q}\neq0$, i.e. those which transform non trivially under translations, it is necessary to study representations of the space group -- yet for commensurate orders, one can consider the simpler task of studying irreps of a so-called extended point group 
\cite{venderbos2016symmetry,venderbos2016multi,wagner2023phenomenology}. The extended point group is the point group of an enlarged unit cell produced by the commensurate CDW. This defines a modified translation subgroup $\widetilde{\mathscr{T}}$: the group of all translations that map the enlarged unit cell to itself. Given the new translation subgroup, a new point group, the extended point group $\widetilde{\mathscr{G}}$, is obtained by taking the factor group $\widetilde{\mathscr{G}}=\mathscr{S} / \widetilde{\mathscr{T}}$. The point group $\widetilde{\mathscr{G}}$ is enlarged by the addition of elements of $\mathscr{T}$ no longer part of $\widetilde{\mathscr{T}}$. Extended point groups are denoted as $\mathscr{G}^{\prime\prime...}$, where the number of primes indicates the number of inequivalent translations in $\mathscr{T}$ but not $\widetilde{\mathscr{T}}$. 

The complex numbers $A(\mathbf{Q}_i,\mathbf{r})$ transform as a representation of the crystal symmetries, and so we can use the group theory methods reviewed in the SM to decompose them into linear combinations corresponding to distinct irreps, effectively filtering STM data into a set of images mapping symmetry-distinct components. In brief, the procedure is to construct group matrices $g$ which realise the action of the symmetry operations on the vector of complex numbers $\{A(\mathbf{Q}_i,\mathbf{r})\}$. The multiplicity $n_{\Gamma}$ of the irrep $\Gamma$ inside this reducible representation can be computed using the characters $\chi(g) = \text{Tr}[g]$, along with the characters of the irreducible representations $\chi_\Gamma(g)$, which can be found in standard character tables (see SM Sec. \ref{supp-char}). One finds
\begin{equation}
\label{n_Gamma}
n_{\Gamma}=\frac{1}{|\widetilde{\mathscr{G}}|} \sum_{g \in \widetilde{\mathscr{G}}} \chi_{\Gamma}(g) \chi(g) 
\end{equation}
We can construct a basis for these irreps by computing the projection operators,
\begin{equation}
\label{proj_eq}
\mathcal{P}_{\Gamma}=\frac{1}{|\widetilde{\mathscr{G}}|} \sum_{g \in \widetilde{\mathscr{G}}} \chi_{\Gamma}(g) \, g 
\end{equation}
Solving for the image of these matrices gives linear combinations of the Bragg peaks which correspond to different irreducible representations; these linear combinations $\Phi_\Gamma(\mathbf{r})$ produce a real space map with the symmetry properties of irrep $\Gamma$.

This demonstrates how to decompose STM data into symmetry channels, but it is important to carefully establish the connection between the symmetries in STM data and the symmetries of order parameters in the electronic Hamiltonian. STM directly images the LDOS, given by $\mathrm{LDOS}(\mathbf{r},E) = \sum_\mathbf{k} |\Psi_\mathbf{k}(\mathbf{r})|^2 \delta(E_\mathbf{k} - E)$, where $E_\mathbf{k}$ is the electronic dispersion and $\Psi_\mathbf{k}(\mathbf{r})$ is the wavefunction~\cite{tersoff1983theory, Coleman2015}. In the absence of symmetry breaking, the LDOS transforms as the trivial irreducible representation (irrep) of the full symmetry group $\mathscr{S}$ of the Hamiltonian (SM Sec. \ref{supp-ldos_symm}). When an order parameter, associated with spontaneous symmetry breaking, reduces the symmetry of the system from $\mathscr{S}$ to a subgroup $\mathscr{S}^\prime$, this breaking will manifest in the LDOS -- any irreps found within the LDOS must be trivial with respect to $\mathscr{S}^\prime$ -- thereby revealing which symmetries have been broken. As a consequence, multiple non-trivial irreps may appear when the order parameter breaks several symmetries -- for instance, CDWs which break mirror symmetry in addition to translations will produce irreps associated with both broken symmetries in the LDOS. These arguments establish a direct correspondence between mean-field symmetry breaking, the symmetry of the LDOS, and the appearance of specific irreps in STM data.

We now make these ideas less abstract with the example of $2\times 2$ CDWs in hexagonal systems. Given a parent system with the $C_{6v}$ point group, a quadrupling of the unit cell implies the breaking of three translational symmetries \cite{venderbos2016symmetry}, and so the relevant extended point group is denoted $C^{\prime\prime\prime}_{6v}$. It can be shown that there are four symmetry-distinct CDWs corresponding the irreps $F_1$, $F_2$, $F_3$, and $F_4$; the irreps $F_2$ and $F_3$ are respectively odd under mirror $\sigma_v:(x,y)\rightarrow (x,-y)$ and $C_{2z}:\mathbf{r}\rightarrow -\mathbf{r}$ symmetries, while $F_4$ is odd under both. Each irrep is three-dimensional, i.e. the order parameters have three independent components.

\begin{figure}[t]
\centering
\includegraphics[width = 0.95\columnwidth]{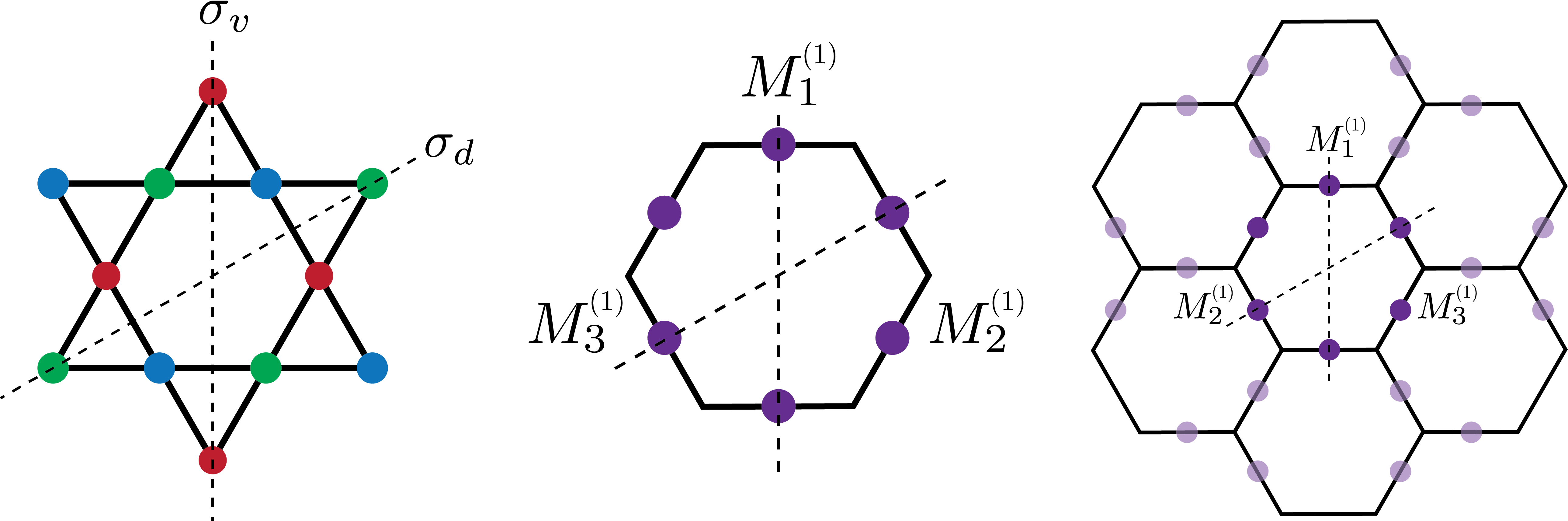}	
\vspace{-0.cm}
\caption{\textbf{Mirror symmetry and its action on the first and higher Brillouin zones.} The kagome lattice possesses the mirror symmetries $\sigma_v$ and $\sigma_d$. The $M$ points of the first Brillouin zone $M^{(1)}_i$ (dark purple) lie on mirror-invariant lines. However, the $M$ points in the second and third Brillouin zones (faint purple) no longer lie on invariant lines, but rather permute among each other under the action of mirror symmetry.}
\label{f:mirror_lines_m}
\vspace{-0.5cm}
\end{figure}

Such a CDW has three distinct wavevectors in the first Brillouin zone, $\mathbf{M}_i$ where $i=1,2,3$, which lie on the midpoints of the zone boundary (Fig. \ref{f:mirror_lines_m}). The STM data can be arranged into a six-dimensional vector $\{A(\mathbf{M}_1,\mathbf{r})$, $A(\mathbf{M}_2,\mathbf{r})$, $A(\mathbf{M}_3,\mathbf{r}),A(-\mathbf{M}_1,\mathbf{r})$, $A(-\mathbf{M}_2,\mathbf{r})$, $A(-\mathbf{M}_3,\mathbf{r})\}$. Since STM data is real-valued, the first and last three components are related by complex conjugation $A(\mathbf{M}_i,\mathbf{r})=A^*(-\mathbf{M}_i,\mathbf{r})$. We construct six-dimensional matrices which act on this vector in a way equivalent to the elements of $C^{\prime\prime\prime}_{6v}$, and then compute the projection operators using the character table for $C^{\prime\prime\prime}_{6v}$ and Eq.~\eqref{proj_eq}. The image space of the projection operators yields linear combinations of the Bragg peaks that isolate each irrep; we find the decomposition (see SM Sec. \ref{supp-proj})
\begin{align}
\label{main:BZ1_decomp}
    \Phi_{F_1}(\mathbf{r}) \!=\! \tfrac{1}{2}\left\{\text{Re} {A}(\mathbf M_1, \mathbf r), \text{Re} {A}(\mathbf M_2, \mathbf r), \text{Re} {A}(\mathbf M_3, \mathbf r)\right\} \nonumber\\
    \Phi_{F_3}(\mathbf{r}) \!=\! \frac{i}{2}\left\{\text{Im} {A}(\mathbf M_1, \mathbf r), \text{Im} {A}(\mathbf M_2, \mathbf r), \text{Im} {A}(\mathbf M_3, \mathbf r)\right\}
\end{align}
while $\Phi_{F_2}(\mathbf{r})=\Phi_{F_4}(\mathbf{r})=0$.

\subsection{Bragg peak extinctions}
The example above reveals a key insight: not all symmetry breaking patterns are detectable in the first Brillouin zone. Consistent with Eq. \eqref{main:BZ1_decomp}, the multiplicity in Eq. \eqref{n_Gamma} yields $n_{\Gamma}=0$ for all $\Gamma$ except $F_1$ and $F_3$, for which $n_{F_1}=n_{F_3}=1$. Hence, information about $F_2$ and $F_4$ \textit{cannot be encoded in the wavevectors of the first Brillouin zone}. In general, a symmetry breaking irrep $\Gamma$ can not be detected from a set of Bragg peaks when Eq. \eqref{n_Gamma} yields $n_\Gamma = 0$; we term this to be a \textit{Bragg peak extinction}. 

\begin{figure*}[t]
\centering
\includegraphics[width = \textwidth]{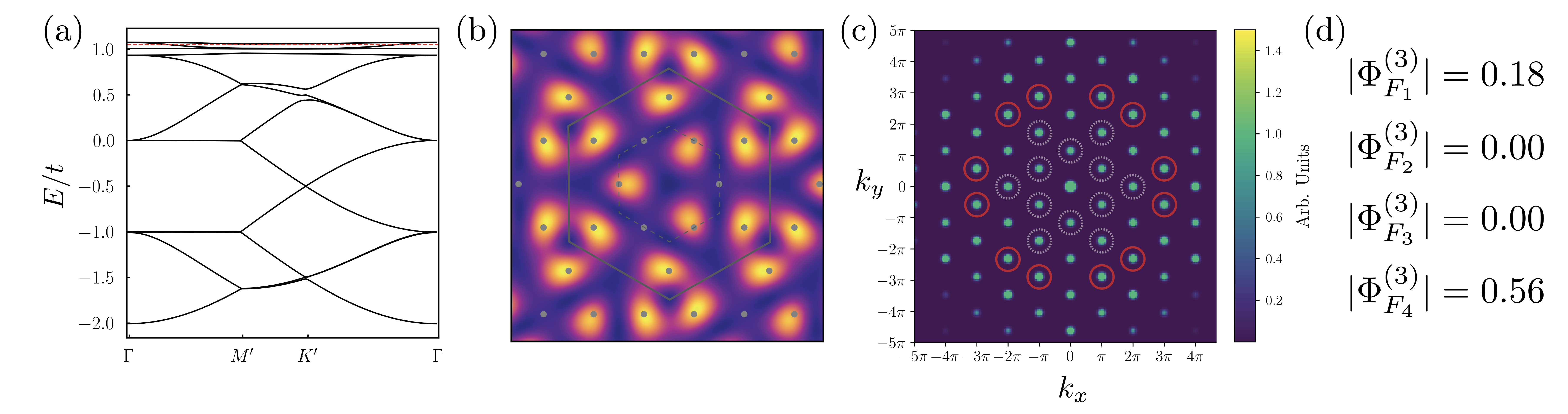}	
\vspace{-0.9cm}
\caption{\textbf{Analysis of synthetic STM data for $F_4$ loop current order.} (a) Bandstructure of the kagome tight-binding model with $F_4$ flux order $\Delta/t=0.1$. (b) $\text{LDOS}(E=2.1, \mathbf{r})$ simulated from the tight-binding model; sites of the kagome lattice are shown in grey, the parent unit cell is shown as a dashed hexagon and the CDW unit cell as a solid hexagon. (c) FFT of the STM data; the Bragg peaks of the CDW in the first and second Brillouin zones are indicated with dashed grey circles, while those in the third zone with solid red. (d) The output of the irrep decomposition acting on the $M$-points of the third Brillouin zone yields a mixture of $F_1$ and $F_4$ symmetries, as predicted by our theory arguments.}
\label{f:synth_main}
\vspace{-0.2cm}
\end{figure*}
This phenomenon has a simple origin. The $F$ irreps are three-dimensional, yet there are only six Bragg peaks in the first Brillouin zone. Imaging four three-component representations requires 12 linearly independent basis functions. When the number of Bragg peaks is less than $12$, it follows that they do not contain enough information to be arranged into 12 linearly independent objects, and therefore cannot image all possible irreps.

The extinction of mirror-odd irreps at the $M$-points of the first Brillouin zone can also be visualised geometrically (Fig. \ref{f:mirror_lines_m}): the six Bragg peaks $\pm \mathbf{M}_i$ lie on three distinct mirror lines, $\sigma_v$, $C_3\sigma_v$, and $C^2_3\sigma_v$. Any contribution to the LDOS that is odd under mirror symmetry must vanish along these lines and therefore at these Bragg peaks.

However, this limitation can be overcome. Moving to higher Brillouin zones, one finds Bragg peaks which transform among each other under these mirror operations, rather than into themselves, indicating that these higher Bragg peaks can encode information about the mirror symmetry breaking irreps $F_2$ and $F_4$. In SM Sec.~\ref{supp-proj}, we present an exhaustive symmetry decomposition in the first, second, and third Brillouin zones for the lattice Bragg peaks, as well as $2\times 2$ and $\sqrt{3}\times \sqrt{3}$ orders, identifying the full set of extinctions. To fully image all possible symmetry breaking patterns, it is necessary to use a larger reducible representation -- either to use Bragg peaks from the second or even third Brillouin zone, or work with a real space basis involving the evaluation of STM data on multiple distinct points within the CDW unit cell (see SM Sec. \ref{supp-real_space}). Physically, both of these methods involve imaging the \textit{intra-unit cell} structure of the CDW -- i.e., certain order parameters can only be symmetry-resolved with intra-unit cell information.

We emphasise that the extinction does \textit{not} imply STM data for a system with $F_{2,4}$ order will be missing Bragg peaks in the first Brillouin zone. To clarify this subtlety, consider perturbing a Hamiltonian with an $F_{4}$ order parameter, breaking translations and vertical mirrors. We established above that the condition for an irrep to appear in the LDOS was that it transforms trivially under the new symmetry group; in addition to $F_4$, now the irreps $F_1$ (broken translations) and $B_2$ (broken mirror) transform trivially -- the $F_1$ signal appears in the first Brillouin zone, while the higher Brillouin zones contain a mixture of $F_1$ and $F_4$. That is, an $F_i$ order parameter induces $F_1$ and $F_i$ LDOS.
The accurate interpretation of the extinction rule is that \textit{evidence of mirror symmetry breaking cannot be detected from the first Brillouin zone peaks}, which will purely comprise the $F_1$ irrep in the presence of an $F_{2,4}$ order.  Hence, STM data does not transform identically to an order parameter -- rather, by imaging all irreps, one can infer which symmetries have been broken.

\begin{figure*}[t]
\centering
\includegraphics[width = \textwidth]{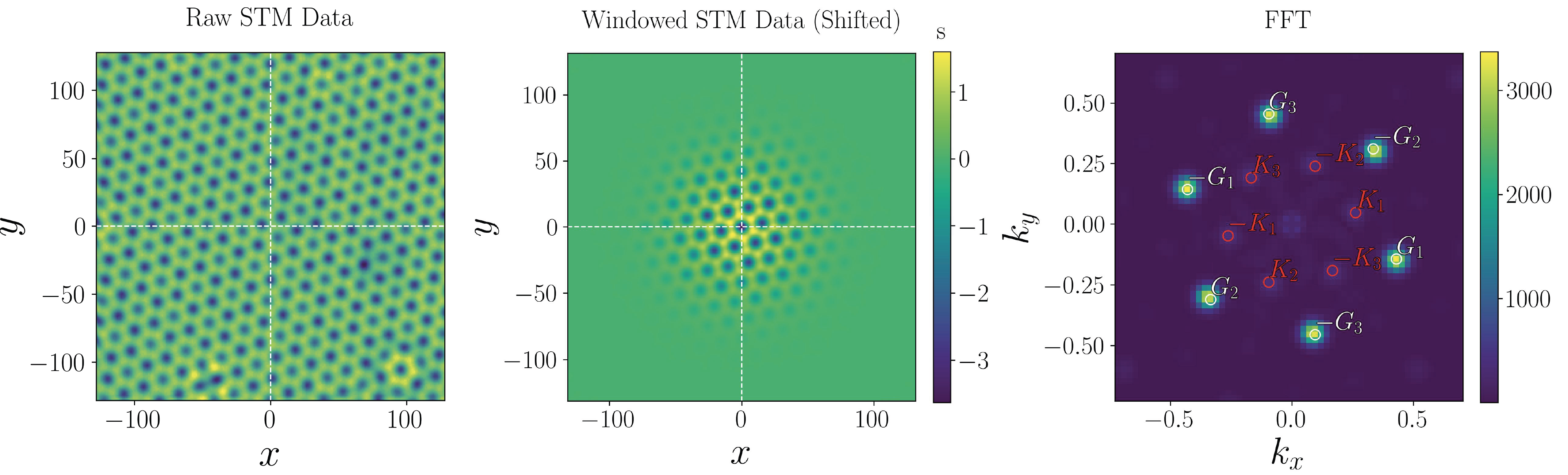}	
\vspace{-0.9cm}
\caption{\textbf{Symmetry analysis of ScV$_6$Sn$_6$ topography.} 
(a) Raw STM data on ScV$_6$Sn$_6$, measured on a $(256,256)$ grid. 
(b) Data after shift correction, $C_6$ symmetrisation and windowing. 
(c) FFT of the processed data, used to extract symmetry components via irrep decomposition. 
This workflow enables quantitative assessment of symmetry leakage and validates the resolution benchmarks reported in the main text.}
\label{f:svs_main}
\vspace{-0.2cm}
\end{figure*}

\subsection{Analysis of synthetic data}

To validate our approach, we now apply the symmetry decomposition to synthetic STM data, which we generate from a tight-binding model of $2\times 2$ CDW order on the kagome lattice. Experiments and theory have suggested the existence of TRS-breaking loop current states in certain kagome metals \cite{Jiang2021, xu2022three, Guo2022switchable, gui2025probing, Shumiya2021, zhou2022chern, christensen2021theory, christensen2022loop, Scammell2023, ingham2025vestigial, ingham2024theory, tazai2022mechanism, tazai2023charge, tazai2024drastic, tazai2024quantum, li2024intertwined}; such states necessarily break mirror symmetries and correspond to the $F_2$ and $F_4$ irreps, and $A_2$ and $B_2$ when the loop currents do not enlarge the unit cell \cite{wagner2023phenomenology}. As shown above, all such mirror-odd irreps are extinct in the first Brillouin zone; the ability to reconstruct them via higher-zone decomposition is therefore a key test of our method.

We consider a model in which the kagome lattice is perturbed by CDW orders of different irreps $F_i$. For illustration, we will consider a single order parameter, corresponding to a time-reversal symmetry breaking $F_4$ irrep, which corresponds to a mirror symmetry breaking flux order; we present results for other cases in the SM. The Hamiltonian takes the form $\mathcal{H} = \mathcal{H}_0+\mathcal{H}_{F_4}$, where 
\begin{align}
    \mathcal{H}_0 = -t\sum_{\langle \mathbf{r}\mathbf{r}'\rangle} c^\dag_{\mathbf{r}} c_{\mathbf{r}'} 
\end{align}
and 
\begin{align}
    \mathcal{H}_{F_4} = i\Delta\sum_{ij,abc} \epsilon_{abc} \cos(\mathbf{M}_a\cdot (\mathbf{R}_i-\mathbf{R}_j)) \,c^\dag_{\mathbf{R}_i,b} c_{\mathbf{R}_j,c}
\end{align}
where $c^\dag_{\mathbf{R}_i,\sigma}$ denotes the electron creation operator at unit cell $\mathbf{R}_i$,  $\sigma=1,2,3$ indexes the three sublattices, while $\langle \mathbf{r} \mathbf{r}'  \rangle$ denote a combination of unit cell and sublattice such that the sites are nearest neighbours. For further details and other order parameter structures see SM Sec. \ref{supp-synth-stm}.

In Fig. \ref{f:synth_main}, we plot the band structure, LDOS, and its Fourier transform, featuring Bragg peaks corresponding to the reciprocal lattice vectors of the parent lattice and the CDW. As discussed in the SM, the $M$-point Bragg peaks of the third Brillouin zone (circled in solid white) can be arranged into maps $\Phi^{(3)}_{F_i}(\mathbf{r})$ that capture all four irreps. We find the non zero irreps to be $|\Phi^{(3)}_{F_1}|\approx 0.18 $ and $|\Phi^{(3)}_{F_1}|\approx 0.56 $. In our analysis of synthetic data for other CDWs, presented in the SM, we confirm our arguments of the previous section: that the LDOS generated from a mean-field order in irrep $F_i$ yields precisely $F_i$ and $F_1$ components, for all $i$.

To correctly filter Fourier-space STM data into irreps, it is essential to accurately determine the relative phases and amplitudes of the Bragg peaks. In addition to the group-theoretic analysis presented in the main text, we outline practical considerations for processing STM data in a way that preserves symmetry information. Fourier decomposition is sensitive to several artefacts:  
($i$) truncation of the topographic image with sharp rectangular boundaries,  
($ii$) experimental imperfections such as spatial drift, and  
($iii$) misalignment of the Fourier origin relative to the unit cell centre across the field of view.  

Among these, artefacts from ($i$) are particularly severe; they induce nematic distortions that lead to spurious leakage into symmetry channels (irreps) absent from the true underlying order. To isolate the effects of ($i$), we analyse synthetic STM data with and without windowing. As shown below and detailed in SM Sec.~\ref{sec:window_synth}, applying a smooth, rotationally symmetric window function [i.e. $w(\mathbf{r})$ in Eq.~\eqref{main:stm}] effectively suppresses these boundary-induced artefacts. This step is critical for obtaining clean and reliable symmetry resolution.

\subsection{Analysis of ScV$_6$Sn$_6$ topography}

We now apply this data preparation protocol to real STM topographic data on ScV$_6$Sn$_6$ (Fig.~\ref{f:svs_main}), a kagome metal from the 166 family that hosts a $\sqrt{3}\times\sqrt{3}$ CDW order~\cite{arachchige2022charge, hu2025flat, gu2023phonon, korshunov2023softening, kundu2024low, jiang2024van, meier2023tiny}. Our goal here is not to determine the CDW symmetry itself, but to demonstrate how real experimental data can be made suitable for symmetry decomposition.

To assess the impact of finite resolution and boundary effects, we construct a $C_6$-symmetrised version of the dataset. Since the allowed irreps under $C_6$ symmetry are known, any spectral weight in forbidden channels provides a direct measure of artefacts introduced by processing. In particular, we evaluate the leakage ratio $|\bm \Phi_G|/|\bm \Phi_{E_1^{'}}|$, which should vanish in perfectly $C_6$-symmetric data.

For the raw STM data on a $(256,256)$ grid, applying a window function alone yields a leakage ratio of $|\bm \Phi_G|/|\bm \Phi_{E_1^{'}}| = 0.31$. In contrast, the same dataset—after $C_6$ symmetrisation and windowing—gives a much smaller leakage of $0.02$. We attribute this residual value to small misalignment in locating the unit cell centre during Fourier processing. Crucially, this comparison confirms that a genuine $|\bm \Phi_G|$ component is present in the data.

These results establish a protocol for reliable symmetry decomposition: apply a rotationally symmetric window function, and verify that irrep leakage is minimised in a symmetrised control dataset. This provides confidence that symmetry components observed in the raw data reflect true physical features rather than artefacts.

\section{Discussion}

Our symmetry decomposition framework enables the direct extraction of subtle point-group symmetry breaking in STM data. By mapping the STM signal onto irreducible representations (irreps) of the extended point group, our method provides a sharp diagnostic of the underlying electronic order, distinguishing among competing or coexisting order parameters. This is particularly valuable for interpreting correlated phases in systems where the symmetry of the order is not obvious.

Our analysis of synthetic data demonstrates that the method can also quantify the relative strength of coexisting order parameters with different symmetry content. This has significant implications for systems with intertwined or composite charge orders, such as moiré materials, as well as kagome metals \cite{hossain2025field}, iron-based superconductors, and the cuprates. In such systems, different irreps may coexist with varying spatial profiles, and symmetry decomposition offers a route to disentangle them.

An important anticipated application lies in the recently studied kagome superconductors $A$V$_3$Sb$_5$ ($A$=K, Rb, Cs), where an unconventional $2\times 2$ charge density wave (CDW) emerges below $T_{\mathrm{CDW}} \approx 90$ K \cite{Neupert2021, Yin2022topological, wilson2024kagome}. STM studies of these materials under applied magnetic fields have reported signatures consistent with CDWs that break time-reversal and mirror symmetries \cite{Jiang2021}; our formalism provides a systematic method to test such symmetry breaking claims and identify the active irreps, promising to resolve the ongoing debate over the symmetries of the CDW. More generally, reanalysing existing STM data on known CDW states using our new methods may shed new light on well-studied materials.

Recent work on 1H transition metal dichalcogenides has also shown that STM can access the topology of underlying bands via real-space probes of symmetry \cite{holbrook2024real, cualuguaru2025probing}. Symmetry decomposition may offer a complementary tool in such settings, resolving not only the presence of broken symmetries but their spatial distribution.

Beyond STM, our approach can be extended to other momentum-resolved probes that couple to order parameters. In particular, neutron scattering studies of magnetically ordered systems could use symmetry decomposition to reveal signatures of loop current order or altermagnetism \cite{liege2024search}. Related ideas have been invoked to study the symmetry of charge order in x-ray studies of cuprate superconductors \cite{comin2015symmetry}; the extension of our formulae to x-ray scattering in a more general setting is another interesting possibility.

More broadly, spatially resolved decomposition into irreps enables a new kind of symmetry microscopy: one that is not limited to detecting order, but can track its spatial structure, quantify coexisting orders, and reveal fluctuations near phase boundaries. This opens new avenues for characterising complex symmetry breaking phases in strongly correlated materials.

\section*{Acknowledgements}
We thank Madisen Holbrook, Dani Muñoz-Segovia, and Raquel Queiroz for discussions.

\bibliography{refs}

\newpage \widetext \newpage

\begin{center}
\textbf{\large Supplementary Material}
\end{center}

\vspace{-0.5cm}
\tableofcontents

\setcounter{equation}{0}
\setcounter{table}{0}
\setcounter{section}{0}
\setcounter{figure}{0}
\makeatletter
\renewcommand{\theequation}{S\arabic{equation}}
\renewcommand{\thefigure}{S\arabic{figure}}
\renewcommand{\thesection}{S\arabic{section}}

\section{Introduction to the Supplementary Materials}
\label{supp-intro}

The Supplementary Materials (SM) present a self-contained guide to our symmetry decomposition method, including theoretical background, implementation procedures, and validation using both synthetic and real STM data.

Here, we provide a high-level summary of each section, highlighting practical takeaways for applying our approach to STM datasets.

\vspace{0.5em}

\noindent
\textbf{Section~\ref{supp-rep}: Representation theory primer.}  
This section introduces key concepts from group theory, including representations and irreducible representations (irreps), with emphasis on how symmetry breaking manifests in the charge density. It serves as a primer for understanding the decomposition formalism.

\vspace{0.5em}

\noindent
\textbf{Section~\ref{supp-char}: Character tables for the extended point groups.}  
We present the character tables required to decompose $2 \times 2$ and $\sqrt{3} \times \sqrt{3}$ CDWs on hexagonal lattices. These tables underpin the projection formula used to extract symmetry-resolved components of the STM signal.

\vspace{0.5em}

\noindent
\textbf{Section~\ref{supp-ldos_symm}: Symmetry properties of the LDOS.}  
Here we discuss how symmetry breaking in the mean-field Hamiltonian affects the local density of states (LDOS). This provides the theoretical justification for interpreting STM signals through symmetry decomposition.

\vspace{0.5em}

\noindent
\textbf{Section~\ref{supp-proj}: Symmetry decomposition of Bragg peaks via the projection method.}  
This section details our main method, in which STM data are decomposed by isolating Bragg peaks in momentum space and constructing appropriate linear combinations, explaining why certain irreps are absent in the first Brillouin zone. Real-space maps of symmetry-distinct components are obtained via inverse Fourier transform. The projection operators are computed using the character tables from Section~\ref{supp-char}.

\vspace{0.5em}

\noindent
\textbf{Section~\ref{supp-convolution}: Symmetry convolution technique.}  
We describe a method to generate synthetic STM maps with controlled symmetry properties by combining real-space basis functions transforming under desired irreps. This technique is useful for testing the sensitivity and selectivity of the decomposition pipeline.

\vspace{0.5em}

\noindent
\textbf{Section~\ref{supp-window}: Window function.}  
This section examines how sharp image boundaries in STM data cause symmetry leakage and nematic artefacts. We show that applying smooth, rotationally symmetric window functions significantly reduces this effect and enhances irrep discrimination. Guidelines for windowing practice are provided.

\vspace{0.5em}

\noindent
\textbf{Section~\ref{supp-real_space}: Real space decomposition.}  
We present the application of real-space irrep decomposition to a kagome lattice with $2 \times 2$ charge order. The method involves sampling at symmetry-distinct intra-unit-cell sites and combining the data according to projection-derived weights. This complements the Bragg peak method and resolves otherwise extinct irreps.

\vspace{0.5em}

\noindent
\textbf{Section~\ref{supp-synth-stm}: Calculation of synthetic STM data.}  
We provide a full description of how synthetic STM maps are generated from tight-binding models. This includes modelling spatial modulations, handling gauge subtleties, and introducing various symmetry breaking terms. We benchmark the decomposition on synthetic maps containing known irreps and assess method accuracy across projection, real-space, and convolution techniques.

\vspace{0.5em}

\noindent
\textbf{Section~\ref{supp-scv6sn6}: Demonstration on ScV$_6$Sn$_6$ topography.}  
The final section applies our symmetry decomposition to real STM data on ScV$_6$Sn$_6$. We assess the effects of windowing and sampling density, and demonstrate that the decomposition converges upon doubling the grid resolution. This serves as a practical case study for preparing real datasets and ensuring artefact suppression.

\section{Representation theory primer}
\label{supp-rep}

To make the formalism employed in this paper more accessible, in this appendix we concisely explain the necessary background in group representation theory. While these ideas are presented in many textbooks, e.g. Ref. \cite{dresselhaus2007group}, this short appendix elucidates only the minimal ingredients to understand our symmetry decomposition approach.

A representation of a group $G$ is a set of square, non-singular matrices $\mathcal{R}(g)$ associated to the elements of a group $g \in G$, such that if $g_{1} g_{2}=g_{3}$ then $\mathcal{R}\left(g_{1}\right) \mathcal{R}\left(g_{2}\right)=\mathcal{R}\left(g_{3}\right)$. In other words, while a group is an abstract collection of objects, a representation is a collection of matrices for which ordinary matrix multiplication respects the group multiplication relation between the group elements. All groups possess a trivial representation in which $\mathcal{R}(g)=1$ for all $g \in G$; by contrast, a representation is said to be faithful when all $\mathcal{R}(g)$ are distinct.

A representation is  reducible if it can be brought to block-diagonal form, 
\begin{align}
\mathcal{R}(g) =\left(\begin{array}{cc}
\mathcal{R}^{(1)}(g) & \mathbf{0} \\
\mathbf{0} & \mathcal{R}^{(2)}(g)
\end{array}\right)
\end{align}
for all $g \in G$. A representation which is not reducible is said to be  irreducible. A reducible representation can always be decomposed into a set of irreducible representations via group theoretic methods we present below. In particular, we concern ourselves with objects which transform linearly under the action of a symmetry operator $\mathcal{O}_g$, known as ``partner functions'' of irrep $\alpha$, denoted $\chi^{\alpha}_i$,
\begin{align}
\mathcal{O}_g \chi^{\alpha}_i = \sum_{ij} \mathcal{R}_{ij}^{(\alpha)}(g)\chi^{\alpha}_j
\end{align}
i.e. objects which transform as vectors acted upon by the matrix representations of the symmetry operators.

The elements of a group can be organised into classes: two elements $g_1$ and $g_2$ belong to the same class if there exists a group element $g$ such that $g_1 = g^{-1} g_2 g$. We also define the so-called characters of a representation --- the traces of the associated matrices, $\chi(g)=\text{Tr}\, \mathcal{R}(g)$. Since $\text{Tr}[g_1]=\text{Tr}[g^{-1}g_1 g]$, matrices belonging to the same class possess the same characters. A character table for a group contains the set of distinct classes, and the characters (columns) associated with each possible irreducible representation (rows). Our motivation for introducing these tables is that the classes and characters can be used to decompose into irreps, as we shall now describe. 

The starting point is the Schur orthogonality theorem, sometimes referred to as the `Grand Orthogonality Theorem'. Denoting the inequivalent unitary irreducible representations of $G$ by $D^{(\alpha)}$, where $\alpha=1, \cdots, n_{r}$, the theorem states
\begin{align}
\sum_{g \in G} \mathcal{R}_{m n}^{(\alpha)}(g)^{*} \mathcal{R}_{m^{\prime} n^{\prime}}^{\left(\alpha^{\prime}\right)}(g)=\frac{|G|}{d_{\alpha}} \delta_{\alpha \alpha^{\prime}} \delta_{m m^{\prime}} \delta_{n n^{\prime}}
\end{align}
where $|G|$ cardinality of $G$ and $d_{\alpha}$ is the dimension of the representation $\mathcal{R}^{(\alpha)}$ \cite{dresselhaus2007group}. 

The orthogonality theorem allows us to construct projection operators which, acting on a given partner function, select out the component of a particular irrep. The projection operator $P^{(\alpha)}$ for an irreducible representation $\alpha$ is given by
\begin{align}
P^{(\alpha)} = \frac{1}{|G|}\sum_{g\in G} \chi^{(\alpha)}(g)\, \mathcal{O}_g
\end{align}
To see that this acts as a projector, consider the action of this operator on a partner function of the irreducible representation $\alpha'$,
\begin{align}
P^{(\alpha')}\chi^{(\alpha')}_i = \frac{1}{|G|}\sum_{g\in G} \chi^{(\alpha)}(g) \,\mathcal{R}_{ij}^{(\alpha')}(g) \,\chi^{(\alpha')}_j 
\end{align}
which, via the Schur orthogonality theorem, vanishes unless $\alpha=\alpha'$. Concretely, we now have a procedure for decomposing objects into irreps: (a) we construct $\mathcal{O}_g$ in the basis of basis functions we wish to decompose, which we can via knowledge of how the symmetries act on our basis functions, and then (b) construct the projection operators using the characters and group elements in each class --- the latter of which can be found in the character table. Having obtained the projection operators, we find which ones have non empty images, and then find a basis for their images. The sections to follow will provide concrete examples.

We note that by taking the trace of the projection operator for irrep $\alpha$ gives a formula for the multiplicity with which a representation appears in this decomposition,
\begin{align}
n_\alpha = \frac{1}{|G|}\sum_{g\in G} \chi^{(\alpha)}(g) \text{Tr} \,\mathcal{O}_g
\end{align}

\newpage

\section{Character tables for the extended point groups}\label{supp-char}

In this appendix, we present the character tables for the extended point groups $C^{\prime\prime}_{6v}$ and $C^{\prime\prime\prime}_{6v}$, respectively relevant to $\sqrt{3}\times\sqrt{3}$ and $2\times 2$ charge density waves in systems with $C_{6v}$ symmetry. \\ \\ \\

\begin{table}[bh!]

    \centering
        \setlength{\tabcolsep}{9pt}
        
    \begin{ruledtabular}
        \begin{tabular}{l|rrrrrrrrrr}
        & & & & & & & & & &  \\
        $\Gamma$ & $\mathcal{C}^{\prime \prime}_1$ & $\mathcal{C}^{\prime \prime}_2$  & $\mathcal{C}^{\prime \prime}_3$  & $\mathcal{C}^{\prime \prime}_4$  & $\mathcal{C}^{\prime \prime}_5$  & $\mathcal{C}^{\prime \prime}_6$  & $\mathcal{C}^{\prime \prime}_7$  & $\mathcal{C}^{\prime \prime}_8$  & $\mathcal{C}^{\prime \prime}_9$  & $\mathcal{C}^{\prime \prime}_{10}$ \\ 
        & & & & & & & & & &  \\  
        \hline 
        & & & & & & & & & &  \\  
         $A_1$ & $1$& $1$ & $1$ & $1$  & $1$ & $1$  &$1$   &$1$ & $1$ &  $1$ \\   
         $A_2$ & $1$& $1$ & $1$ & $1$  & $1$ & $1$  &$- 1$ & $-1$ & $-1$ & $-1$ \\  
         $B_1$ & $1$& $1$ & $-1$& $-1$ & $1$ & $-1$ &$ 1$  & $1$ & $-1$ & $-1$  \\  
         $B_2$ & $1$& $1$ & $-1$& $-1$ & $1$ & $-1$ &$- 1$ & $-1$ & $1$ & $1$ \\  
         $E_1$ & $2$& $2$ & $-2$& $-2$ & $-1$& $1$  &$ 0$  & $0$ & $0$ &  $0$ \\  
         $E_2$ & $2$& $2$ & $2$ & $2$  & $-1$& $-1$ &$ 0$  & $0$ & $0$ &  $0$ \\ 
        & & & & & & & & & &  \\
        \hline
        & & & & & & & & & &  \\  
         $F_1$ & $3$& $-1$& $3$ & $-1$ & $0$& $0$  &$ 1$  & $-1$ & $1$ & $-1$  \\  
         $F_2$ & $3$& $-1$& $3$ & $-1$ & $0$& $0$  &$ -1$ & $1$ & $-1$ &  $ 1$ \\  
         $F_3$ & $3$& $-1$& $-3$& $1$  & $0$ & $0$  &$ 1$  & $-1$ & $-1$ &  $1$  \\  
         $F_4$ & $3$& $-1$& $-3$& $1$  & $0$ & $0$  &$ -1$ & $1$ & $1$ & $-1$ \\[0.2 cm]  
        \end{tabular}
    \end{ruledtabular}
    
    \caption[Character table for extended point group $C_{6v}^{\prime\prime\prime}$.]{The character table for the point group $C^{\prime\prime\prime}_{6v}$, where $t_1$, $t_2$, $t_3$ correspond to $T(\mathbf{a}_1)$, $T(\mathbf{a}_2)$, and $T(\mathbf{a}_1+\mathbf{a}_2)$. The irreducible representations which arise as a consequence of the added translations are $F_1$, $F_2$, $F_3$ and $F_4$, all three-dimensional. The group operations contained in each conjugacy class are $\mathcal{C}_1^{\prime\prime\prime}=\{I\}$, $\mathcal{C}_2^{\prime\prime\prime} = \{t_i\}$, $\mathcal{C}_3^{\prime\prime\prime}=\{C_2\}$, $\mathcal{C}_4^{\prime\prime\prime} = \{t_iC_2\}$, $\mathcal{C}_5^{\prime\prime\prime} = \{C_3, C_3^{-1}, t_iC_3, t_iC_3^{-1}\}$, $\mathcal{C}_6^{\prime\prime\prime} = \{C_6, C_6^{-1}, t_i C_6, t_i C_6^{-1}\}$, $\mathcal{C}_7^{\prime\prime\prime} = \{\sigma_{vi}, t_i\sigma_{vi}\}$, $\mathcal{C}_8^{\prime\prime\prime} = \{t_i \sigma_{vj}\}$, $\mathcal{C}_9^{\prime\prime\prime} = \{\sigma_{di}, t_i\sigma_{di}\}$, $\mathcal{C}_{10}^{\prime\prime\prime} = \{t_i\sigma_{dj}\}$. See Ref. \cite{wagner2023phenomenology}.}
    \label{tab:cppp6v}
\end{table}

\begin{table}[bh!]
    \centering
        \setlength{\tabcolsep}{9pt}
        
    \begin{ruledtabular}
        \begin{tabular}{l|rrrrrrrrr}
        & & & & & & & & &  \\
        $\Gamma$ & $\mathcal{C}^{\prime \prime}_1$ & $\mathcal{C}^{\prime \prime}_2$  & $\mathcal{C}^{\prime \prime}_3$  & $\mathcal{C}^{\prime \prime}_4$  & $\mathcal{C}^{\prime \prime}_5$  & $\mathcal{C}^{\prime \prime}_6$  & $\mathcal{C}^{\prime \prime}_7$ & $\mathcal{C}^{\prime \prime}_8$ & $\mathcal{C}^{\prime \prime}_9$\\ 
        & & & & & & & & & \\
        \hline 
        & & & & & & & & & \\
        $A_1$ & 1 & 1 & 1 & 1 & 1 & 1 & 1 & 1 & 1 \\
        $A_2$ & 1 & 1 & 1 & 1 & 1 & 1 & $-1$ & $-1$ & $-1$ \\
        $B_2$ & 1 & 1 & $-1$ & 1 & 1 & $-1$ & $-1$ & $-1$ & 1 \\
        $B_1$ & 1 & 1 & $-1$ & 1 & 1 & $-1$ & 1 & 1 & $-1$ \\
        $E_1$ & 2 & 2 & $-2$ & $-1$ & $-1$ & 1 & 0 & 0 & 0 \\
        $E_2$ & 2 & 2 & 2 & $-1$ & $-1$ & $-1$ & 0 & 0 & 0 \\
        & & & & & & & & &  \\  
        \hline
        & & & & & & & & &  \\  
        $E_1^{\prime}$ & 2 & $-1$ & 0 & 2 & $-1$ & 0 & 2 & $-1$ & 0 \\
        $E_2^{\prime}$ & 2 & $-1$ & 0 & 2 & $-1$ & 0 & $-2$ & 1 & 0 \\
        $G^{\prime}$ & 4 & $-2$ & 0 & $-2$ & 1 & 0 & 0 & 0 & 0 \\[0.2 cm] 
        \end{tabular} 
    \end{ruledtabular}
    
    \caption[Character table for extended point group $C_{6v}^{\prime\prime}$.]{The character table for the point group $C''_{6v}$. The irreducible representations which arise as a consequence of the added translations are $E'_1$ and $E'_2$, which are two-dimensional, as well as $G'$, which is four-dimensional. The group operations in each conjugacy class are $\mathcal{C}_1^{\prime\prime}=I$, $\mathcal{C}_2^{\prime\prime} = t_i$, $\mathcal{C}_3^{\prime\prime}= C_2, t_iC_2$, $\mathcal{C}_4^{\prime\prime} = C_3, C_3^{-1}$, $\mathcal{C}_5^{\prime\prime} = t_iC_3, t_iC_3^{-1}$, $\mathcal{C}_6^{\prime\prime} = C_6, C_6^{-1}, t_i C_6, t_i C_6^{-1}$, $\mathcal{C}_7^{\prime\prime} = \sigma_{di}$, $\mathcal{C}_8^{\prime\prime} = t_i \sigma_{di}$, $\mathcal{C}_9^{\prime\prime} =\sigma_{vi},t_i\sigma_{vi}$. See Ref. \cite{basko2008theory}.}
    \label{tab:cpp6v}
\end{table}

\newpage

\section{Symmetry Properties of the Local Density of States}\label{supp-ldos_symm}

The local density of states (LDOS) at position \( \mathbf{r} \) and energy \( E \) is given by
\begin{align}
\rho(\mathbf{r}, E) = \sum_n |\psi_n(\mathbf{r})|^2 \delta(E - E_n),
\end{align}
where \( \{ \psi_n \} \) are the eigenfunctions of a Hamiltonian \( H \) with eigenvalues \( E_n \). More generally, if some eigenvalues are degenerate, we label the orthonormal eigenfunctions within the degenerate subspace as \( \psi_{n,\alpha} \), where \( \alpha = 1, \dots, d_n \) and \( d_n \) is the degeneracy of level \( E_n \). Then,
\begin{align}
\rho(\mathbf{r}, E) = \sum_n \sum_{\alpha=1}^{d_n} |\psi_{n,\alpha}(\mathbf{r})|^2 \delta(E - E_n).
\end{align}
Let \( G \) denote the symmetry group of the Hamiltonian, and let \( g \in G \) be a symmetry operation with corresponding unitary operator \( U_g \) such that \( [U_g, H] = 0 \). Within each degenerate eigenspace, the symmetry acts via a unitary representation \( D^{(n)}(g) \in U(d_n) \), i.e.
\begin{align}
U_g \psi_{n,\alpha} = \sum_{\beta=1}^{d_n} D^{(n)}_{\beta\alpha}(g) \psi_{n,\beta}.
\end{align}
Under \( g \), the point \( \mathbf{r} \) maps to \( g \cdot \mathbf{r} \), and
\begin{align}
\rho(g \cdot \mathbf{r}, E) &= \sum_n \sum_{\alpha=1}^{d_n} |\psi_{n,\alpha}(g \cdot \mathbf{r})|^2 \delta(E - E_n) \nonumber \\
&= \sum_n \sum_{\alpha=1}^{d_n} |(U_g \psi_{n,\alpha})(\mathbf{r})|^2 \delta(E - E_n) \nonumber \\
&= \sum_n \sum_{\alpha=1}^{d_n} \left| \sum_{\beta=1}^{d_n} D^{(n)}_{\beta\alpha}(g) \psi_{n,\beta}(\mathbf{r}) \right|^2 \delta(E - E_n).
\end{align}
Using unitarity of \( D^{(n)}(g) \), we simplify the sum over \( \alpha \):
\begin{align}
\sum_{\alpha=1}^{d_n} \left| \sum_{\beta=1}^{d_n} D^{(n)}_{\beta\alpha}(g) \psi_{n,\beta}(\mathbf{r}) \right|^2 
&= \sum_{\beta=1}^{d_n} |\psi_{n,\beta}(\mathbf{r})|^2.
\end{align}
We therefore arrive at the unsurprising conclusion that the charge density is invariant under all the symmetries of the Hamiltonian,
\begin{align}
\rho(g \cdot \mathbf{r}, E) = \rho(\mathbf{r}, E),
\end{align}

This invariance holds even if the individual eigenfunctions \( \psi_{n,\alpha} \) transform under a non-trivial irrep of \( G \). For instance, in the case of a band corresponding to a higher-dimensional irreducible representation, the sum over squared amplitudes \( \sum_\alpha |\psi_{n,\alpha}(\mathbf{r})|^2 \) is still invariant under the group action due to the unitarity of the representation. Hence, probing the LDOS at the energy of a non-trivial band still yields a scalar quantity that respects the full symmetry of the Hamiltonian.

The fact that the charge density must transform as the trivial (identity) irreducible representation of the symmetry group \( G \) seems like a fairly obvious observation, yet it forms an important premise in our relation of the LDOS to symmetry breaking in the following subsection.

\subsection{LDOS with symmetry breaking}
For illustration, we consider a parent Hamiltonian $H_0$ which possesses the $C_{6v}$ point group symmetry, but our line of reasoning applies more generally. In this case, the LDOS is invariant under all operations of $C_{6v}$, and thus transforms as the trivial irreducible representation $A_1$ of $C_{6v}$.

Now consider a symmetry breaking perturbation that lowers the symmetry of the Hamiltonian to a subgroup $\tilde{G} \subset C_{6v}$. In particular, suppose the perturbation corresponds to an order parameter $\Phi_{B_2}$ transforming as the $B_2$ irrep of $C_{6v}$. This breaks the vertical mirror symmetry $\sigma_v: x \rightarrow -x$.
Then the new Hamiltonian is
\begin{align}
H = H_0 + \delta H_{B_2},
\end{align}
which preserves only the symmetries in $\tilde{G}$, the $B_2$-invariant subgroup of $C_{6v}$. After symmetry breaking, the LDOS must still transform as the trivial irrep of the new symmetry group $\tilde{G}$, i.e. it must be invariant under the residual symmetries. However, this allows contributions that were not invariant under the original $C_{6v}$ group.

To determine which irreps of $C_{6v}$ can now contribute to $\rho(\mathbf{r}, E)$, we consider how irreps restrict to $\tilde{G}$. In particular, we note
\begin{align}
\left. A_1 \right|_{\tilde{G}} &= A_1', \\
\left. B_1 \right|_{\tilde{G}} &= A_1'
\end{align}
That is, both the $A_1$ and $B_1$ irreps of $C_{6v}$ reduce to the trivial representation of the reduced group $\tilde{G}$. Thus, both can appear in the LDOS after the symmetry is lowered:
\begin{align}
\rho(\mathbf{r}, E) = f_{A_1}(\mathbf{r}) + f_{B_1}(\mathbf{r}),
\end{align}
where $f_{A_1}$ and $f_{B_1}$ transform under $C_{6v}$ as $A_1$ and $B_1$, respectively, but are both invariant under $\tilde{G}$.  

By the same arguments, when considering CDW ordering in the $F_i$ irrep, the Hamiltonian
\begin{align}
H = H_0 + \delta H_{F_i},
\end{align}
preserves only the symmetries in $\tilde{G}$, the $F_i$-invariant subgroup of $C_{6 v}^{\prime \prime \prime}$. Then
\begin{align}
\left. F_1 \right|_{\tilde{G}} &= A_1', \\
\left. F_i \right|_{\tilde{G}} &= A_1'
\end{align}
That is, both the $F_1$ and $F_i$ irreps of $C_{6 v}^{\prime \prime \prime}$ reduce to the trivial representation of the reduced group $\tilde{G}$. Hence LDOS always observes an $F_1$ component for any imposed $F_i$ mean field order parameter. In addition, one expects irreps other than $A_1$ to appear in the symmetry decomposition of the lattice Bragg peaks -- for instance, $F_3$ order should result in the reciprocal lattice Bragg peaks exhibiting a $B_1$ contribution.

\newpage

\section{Symmetry decomposition of Bragg peaks via the symmetry projection operators}\label{supp-proj}

In this section we shall present the results for symmetry-decomposition of the Bragg peaks of the lattice as well as $2 \times 2$ and $\sqrt{3} \times \sqrt{3}$ CDWs, in the first Brillouin zone and beyond. We will begin by explicitly demonstrating how the calculation works for the case of $2\times 2$ CDWs.

\subsection{Method: illustration for $2\times 2$ order}

We consider a parent system with $C_{6v}$ symmetry, such as the kagome lattice, and investigate a $2\times 2$ enlargement of the unit cell; such states have been theoretically predicted as naturally arising from electronic correlations \cite{wenger2024theory}, and have been seen for instance in kagome metals $A$V$_3$Sb$_5$, and transition metal dichalcogenide 1T-TiSe$_2$ \cite{di1976electronic, rossnagel2011origin, calandra2011charge, kogar2017signatures}. The $2\times 2$ order implies the breaking of three translational symmetries: $T\left(\mathbf{a}_{1}\right) \equiv t_{1}, T\left(\mathbf{a}_{2}\right) \equiv t_{2}$ and $T\left(\mathbf{a}_{1}+\mathbf{a}_{2}\right) \equiv t_{3}$. The hexagonal group $C_{6 v}$, which has 12 elements, while the group $C_{6 v}^{\prime \prime \prime}$ also contains $t_{1,2,3}$, and as a result consists of 48 elements. The character table of $C_{6 v}^{\prime \prime \prime}$ is given in Table \ref{tab:cppp6v}, c.f. Ref. \cite{venderbos2016symmetry}.

We proceed by calculating how the symmetry operators act on the vector $\{A(\mathbf{M}_1,\mathbf{r})$, $A(\mathbf{M}_2,\mathbf{r})$, $A(\mathbf{M}_3,\mathbf{r})$, $A(-\mathbf{M}_1,\mathbf{r})$, $A(-\mathbf{M}_2,\mathbf{r})$, $A(-\mathbf{M}_3,\mathbf{r}))\}$, as discussed in the main text. Using explicit knowledge of how the vectors $\pm \mathbf{M}_i$ transform among each other under the symmetries, we can write down the matrices which represent the action of the symmetry operators in this basis. One finds the elementary translations
\[
g_{t_1}=\begin{pmatrix}
-1 & 0 & 0 & 0 & 0 & 0 \\
0 & 1 & 0 & 0 & 0 & 0 \\
0 & 0 & -1 & 0 & 0 & 0 \\
0 & 0 & 0 & -1 & 0 & 0 \\
0 & 0 & 0 & 0 & 1 & 0 \\
0 & 0 & 0 & 0 & 0 & -1
\end{pmatrix}, \ \
g_{t_2}=\begin{pmatrix}
-1 & 0 & 0 & 0 & 0 & 0 \\
0 & -1 & 0 & 0 & 0 & 0 \\
0 & 0 & 1 & 0 & 0 & 0 \\
0 & 0 & 0 & -1 & 0 & 0 \\
0 & 0 & 0 & 0 & -1 & 0 \\
0 & 0 & 0 & 0 & 0 & 1
\end{pmatrix}
\]
The rotation operators are represented as
\[
g_{C_2}=\begin{pmatrix}
0 & 0 & 0 & 1 & 0 & 0 \\
0 & 0 & 0 & 0 & 1 & 0 \\
0 & 0 & 0 & 0 & 0 & 1 \\
1 & 0 & 0 & 0 & 0 & 0 \\
0 & 1 & 0 & 0 & 0 & 0 \\
0 & 0 & 1 & 0 & 0 & 0
\end{pmatrix}, \ \
g_{C_3}=\begin{pmatrix}
0 & 1 & 0 & 0 & 0 & 0 \\
0 & 0 & 1 & 0 & 0 & 0 \\
1 & 0 & 0 & 0 & 0 & 0 \\
0 & 0 & 0 & 0 & 1 & 0 \\
0 & 0 & 0 & 0 & 0 & 1 \\
0 & 0 & 0 & 1 & 0 & 0
\end{pmatrix}
\]
Lastly, the vertical and diagonal mirrors are given by
\[
g_{\sigma_v}=\begin{pmatrix}
0 & 0 & 1 & 0 & 0 & 0 \\
0 & 1 & 0 & 0 & 0 & 0 \\
1 & 0 & 0 & 0 & 0 & 0 \\
0 & 0 & 0 & 0 & 0 & 1 \\
0 & 0 & 0 & 0 & 1 & 0 \\
0 & 0 & 0 & 1 & 0 & 0
\end{pmatrix}, \ \
g_{\sigma_d}=\begin{pmatrix}
0 & 0 & 0 & 0 & 0 & 1 \\
0 & 0 & 0 & 0 & 1 & 0 \\
0 & 0 & 0 & 1 & 0 & 0 \\
0 & 0 & 1 & 0 & 0 & 0 \\
0 & 1 & 0 & 0 & 0 & 0 \\
1 & 0 & 0 & 0 & 0 & 0
\end{pmatrix}
\]
All the other elements of the group can be constructed as products of this elementary set.

To arrange the Bragg peaks into irreps of the extended point group, we note that the multiplicity $n_{\Gamma}$ of the irrep $\Gamma$ in the decomposition of this reducible representation can be computed using the characters $\chi(g)$, i.e. the traces of the above matrices,
\begin{equation}
n_{\Gamma}=\frac{1}{\left|C_{6 v}^{\prime \prime \prime}\right|} \sum_{g \in C_{6 v}^{\prime \prime \prime}} \chi_{\Gamma}(g) \chi(g) 
\end{equation}
We find that $n_{\Gamma}=0$ for all $\Gamma$ except $F_1$ and $F_3$, for which $n_{F_1}=n_{F_3}=1$. Both $F$ irreps are three dimensional; the fact we find two three dimensional representations is in agreement with the fact that the Bragg peaks taken together form a six dimensional representation. Next, we can construct a basis for these irreps by computing the projection operators,
\begin{equation}
\mathcal{P}_{\Gamma}=\frac{1}{\left|C_{6 v}^{\prime \prime \prime}\right|} \sum_{g \in C_{6 v}^{\prime \prime \prime}} \chi_{\Gamma}(g) \, g 
\end{equation}
and solving for their image. The result is 
\begin{align}
\label{F1}
\mathbf{\Phi}_{F_{1}}(\mathbf r)&=\frac{1}{2}\left\{\text{Re} {A}(\mathbf M_1, \mathbf r), \text{Re} {A}(\mathbf M_2, \mathbf r), \text{Re} {A}(\mathbf M_3, \mathbf r)\right\}\\
\label{F3}
\mathbf{\Phi}_{F_{3}}(\mathbf r)&=\frac{i}{2}\left\{\text{Im} {A}(\mathbf M_1, \mathbf r), \text{Im} {A}(\mathbf M_2, \mathbf r), \text{Im} {A}(\mathbf M_3, \mathbf r)\right\}
\end{align}
The upshot is that the real and imaginary parts of the Bragg peaks correspond to separate irreducible representations of the symmetry group. In situations where we expect $C_3$ symmetry to be further broken, we can reorganise each $F$ channel into an $A$-like and $E$-like channel, e.g. 
\begin{align}
\mathbf{\Phi}^{A}_{F_{1}}(\mathbf r)&=\frac{1}{\sqrt{3}}\left(\text{Re} {A}(\mathbf M_1, \mathbf r) + \text{Re} {A}(\mathbf M_2, \mathbf r)+\text{Re} {A}(\mathbf M_3, \mathbf r)\right)\\
\mathbf{\Phi}^{E}_{F_{1}}(\mathbf r)&=\frac{1}{\sqrt{6}}\left\{2\text{Re} {A}(\mathbf M_1, \mathbf r)- \text{Re} {A}(\mathbf M_2, \mathbf r)  - \text{Re} {A}(\mathbf M_3, \mathbf r),  \sqrt{3}(\text{Re} {A}(\mathbf M_2, \mathbf r) - \text{Re} {A}(\mathbf M_3, \mathbf r)) \right\}
\end{align}
and similar expressions for $F_3$. The breaking of $C_3$ symmetry is signified by nonzero values for the $\Phi^{E}$ channels, and variation of the magnitude of these two channels indicates a variation in the local residual $C_2$ axis, if one exists.

\subsubsection{Higher-order $2\times 2$ CDW Bragg peaks}
\label{supp-M-bz3}

\begin{figure*}[t]
\centering
\includegraphics[width=\textwidth]{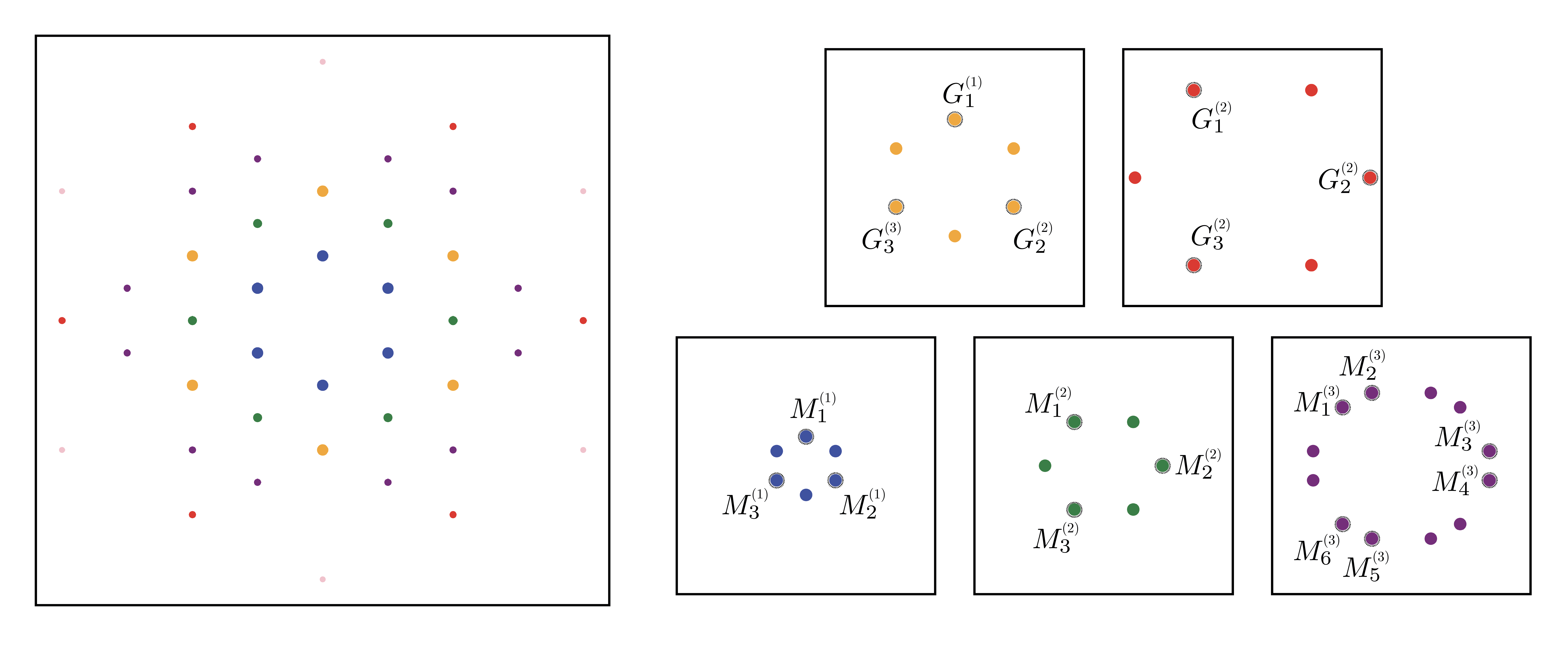}
\vspace{-0.3cm}
\caption{\textbf{Bragg peaks for $2\times 2$ charge density wave.} Left: Bragg peaks in the first three Brillouin zones for a hexagonal lattice with $2\times 2$ CDW order, with the dot size decreasing for wavevectors of larger magnitude. Right: definitions of the wavevectors $\mathbf{G}_i^{(n)}$, $\mathbf{M}_i^{(n)}$ colour-coded to match the left image. Orange and red are the first and second shell of lattice Bragg peaks respectively, while blue, green, and purple are the first, second and third shells of CDW Bragg peaks. Note that in our definitions $\mathbf{G}_1$ points entirely along the $k_y$ direction --- in principle an image of STM data may correspond to a rotation of this picture by some angle depending on the sample orientation.}
\label{f:supp_fig1}
\vspace{0cm}
\end{figure*}

In Fig. \ref{f:supp_fig1}, we illustrate a large set of Bragg peaks in the presence of a $2\times 2$ CDW in the first second, including the lattice Bragg peaks of the lattice. We show the first two shells of lattice Bragg peaks as well as the first three shells of CDW Bragg peaks, which are individually identified on the right. To give explicit expressions, $\{\mathbf{M}^{(2)}_1, \mathbf{M}^{(2)}_2, \mathbf{M}^{(2)}_3\} = \{\mathbf{G}_1+\mathbf{G}_2-\mathbf{M}_3, -\mathbf{G}_2-\mathbf{M}_1, \mathbf{G}_2 + \mathbf{M}_2\}$, and $\{\mathbf{M}^{(3)}_1, \mathbf{M}^{(3)}_2, \mathbf{M}^{(3)}_3, \mathbf{M}^{(3)}_4, \mathbf{M}^{(3)}_5, \mathbf{M}^{(3)}_6 \} = \{\mathbf{G}_1+\mathbf{G}_2+\mathbf{M}_1, \mathbf{G}_1 -\mathbf{M}_2, -\mathbf{G}_1-2\mathbf{G}_2-\mathbf{M}_2,-\mathbf{G}_1-2\mathbf{G}_2+\mathbf{M}_1, -\mathbf{G}_1+\mathbf{M}_3, -\mathbf{G}_1+\mathbf{G}_2+\mathbf{M}_1\}$.

To decompose these Bragg peaks into symmetry channels, one employs the same methods explicitly illustrated in the previous subsection, i.e. construct the representation of the symmetry operators in the basis of these Bragg peaks, construct the projection operators and then solve for their image. We simply state the results rather than present the explicit form of the symmetry representations.

In the second shell we find $F_4$ is now visible 
\begin{align}
\label{F1_2}
\mathbf{\Phi}_{F^{(2)}_{1}}(\mathbf r)&=\frac{1}{2}\left\{\text{Re} {A}(\mathbf{M}^{(2)}_1, \mathbf r), \text{Re} {A}(\mathbf{M}^{(2)}_2, \mathbf r), \text{Re} {A}(\mathbf{M}^{(2)}_3, \mathbf r)\right\}\\
\label{F2_2}
\mathbf{\Phi}_{F^{(2)}_{4}}(\mathbf r)&=\frac{i}{2}\left\{\text{Im} {A}(\mathbf{M}^{(2)}_1, \mathbf r), \text{Im} {A}(\mathbf{M}^{(2)}_2, \mathbf r), \text{Im} {A}(\mathbf{M}^{(2)}_3, \mathbf r)\right\}
\end{align}
As usual one can split these into $A$-like and $E$-like irreps to encode threefold rotational symmetry breaking as well.

In the third shell of CDW Bragg peaks, every irrep becomes visible, with the symmetry decomposition,
\begin{align}
\label{F1_3}
\mathbf{\Phi}_{F^{(3)}_{1}}(\mathbf r)&=\frac{1}{2}\left\{\text{Re} \left[ {A}(\mathbf{M}^{(3)}_1, \mathbf r)+{A}(\mathbf{M}^{(3)}_6, \mathbf r)\right], \text{Re} \left[ {A}(\mathbf{M}^{(3)}_2, \mathbf r)+{A}(\mathbf{M}^{(3)}_3, \mathbf r)\right], \text{Re} \left[ {A}(\mathbf{M}^{(3)}_4, \mathbf r)+{A}(\mathbf{M}^{(3)}_5, \mathbf r)\right]\right\}\\
\label{F2_3}
\mathbf{\Phi}_{F^{(3)}_{2}}(\mathbf r)&=\frac{1}{2}\left\{\text{Re} \left[ {A}(\mathbf{M}^{(3)}_1, \mathbf r)-{A}(\mathbf{M}^{(3)}_6, \mathbf r)\right], \text{Re} \left[ {A}(\mathbf{M}^{(3)}_3, \mathbf r)-{A}(\mathbf{M}^{(3)}_2, \mathbf r)\right], \text{Re} \left[ {A}(\mathbf{M}^{(3)}_5, \mathbf r)-{A}(\mathbf{M}^{(3)}_4, \mathbf r)\right]\right\}\\
\label{F3_3}
\mathbf{\Phi}_{F^{(3)}_{3}}(\mathbf r)&=\frac{i}{2}\left\{\text{Im} \left[ {A}(\mathbf{M}^{(3)}_1, \mathbf r)-{A}(\mathbf{M}^{(3)}_6, \mathbf r)\right], \text{Im} \left[ {A}(\mathbf{M}^{(3)}_3, \mathbf r)-{A}(\mathbf{M}^{(3)}_2, \mathbf r)\right], \text{Im} \left[ {A}(\mathbf{M}^{(3)}_5, \mathbf r)-{A}(\mathbf{M}^{(3)}_4, \mathbf r)\right]\right\}\\
\label{F4_3}
\mathbf{\Phi}_{F^{(3)}_{4}}(\mathbf r)&=\frac{i}{2}\left\{\text{Im} \left[ {A}(\mathbf{M}^{(3)}_1, \mathbf r)+{A}(\mathbf{M}^{(3)}_6, \mathbf r)\right], \text{Im} \left[ {A}(\mathbf{M}^{(3)}_2, \mathbf r)+{A}(\mathbf{M}^{(3)}_3, \mathbf r)\right], \text{Im} \left[ {A}(\mathbf{M}^{(3)}_4, \mathbf r)+{A}(\mathbf{M}^{(3)}_5, \mathbf r)\right]\right\}
\end{align}
Yet again, these basis functions can be combined into $A$-like and $E$-like irreps, using the same combination of coefficients used as in the lower order Bragg peaks, i.e. for an irrep vector $\mathbf{\Phi} = (\Phi_1, \Phi_2, \Phi_3)$, one has $\Phi^A = \tfrac{1}{\sqrt{3}}(\Phi_1+\Phi_2+\Phi_3)$ and $\mathbf{\Phi}^E = \tfrac{1}{\sqrt{6}}(2\Phi_1-\Phi_2-\Phi_3, \sqrt{3}(\Phi_2-\Phi_3))$.

\subsection{Symmetry channels for lattice Bragg peaks}
\label{supp-lattice-bragg}

While the decomposition of the CDW Bragg peaks gives symmetry insight into the nature of the density of states from the CDW, the Bragg peaks give insight into translationally invariant orders. The construction of the symmetry operators for the Bragg peaks of the lattice is very similar, except these Fourier components definitionally transform trivially under all the translation operators rather than pick up minus signs.

One finds that the first shell of lattice Bragg peaks contains the irreps $A_1$, $B_1$, $E_1$ and $E_2$; the one-dimensional basis functions are 
\begin{align}
\label{A1}
\mathbf{\Phi}_{A_{1}}(\mathbf r)&=\frac{1}{\sqrt{3}}\left\{\text{Re} {A}(\mathbf G^{(1)}_1, \mathbf r) + \text{Re} {A}(\mathbf{G}^{(1)}_2, \mathbf r) + \text{Re} {A}(\mathbf{G}^{(1)}_3, \mathbf r)\right\}\\
\label{B1}
\mathbf{\Phi}_{B_1}(\mathbf r)&=\frac{i}{\sqrt{3}}\left\{\text{Im} {A}(\mathbf G^{(1)}_1, \mathbf r) + \text{Im} {A}(\mathbf{G}^{(1)}_2, \mathbf r) + \text{Im} {A}(\mathbf{G}^{(1)}_3, \mathbf r)\right\}
\end{align}
while the two-dimensional irreps are two component vectors
\begin{align}
\label{E1}
\mathbf{\Phi}_{E_1}(\mathbf r)&=\frac{1}{\sqrt{6}}\left\{2\text{Re} {A}(\mathbf G^{(1)}_1, \mathbf r)- \text{Re} {A}(\mathbf{G}^{(1)}_2, \mathbf r)  - \text{Re} {A}(\mathbf{G}^{(1)}_3, \mathbf r), \sqrt{3}(\text{Re} {A}(\mathbf{G}^{(1)}_2, \mathbf r) - \text{Re} {A}(\mathbf{G}^{(1)}_3, \mathbf r)) \right\}\\
\label{E2}
\mathbf{\Phi}_{E_2}(\mathbf r)&=\frac{i}{\sqrt{6}}\left\{2\text{Im} {A}(\mathbf G^{(1)}_1, \mathbf r)- \text{Im} {A}(\mathbf{G}^{(1)}_2, \mathbf r)  - \text{Im} {A}(\mathbf{G}^{(1)}_3, \mathbf r),  \sqrt{3}(\text{Im} {A}(\mathbf{G}^{(1)}_2, \mathbf r) - \text{Im} {A}(\mathbf{G}^{(1)}_3, \mathbf r)) \right\}
\end{align}
The $A_2$ and $B_2$ irreps are invisible in the first Brillouin zone. Note we had six complex numbers and hence only obtain six linearly independent channels.

\subsubsection{Higher-order lattice Bragg peaks}

The explicit expressions for the second shell of Bragg peaks are $\{\mathbf{G}^{(2)}_1, \mathbf{G}^{(2)}_2, \mathbf{G}^{(2)}_3 \} = \{2\mathbf{G}_1 + 2\mathbf{G}_2, -\mathbf{G}_1-2\mathbf{G}_2, -\mathbf{G}_1+\mathbf{G}_2\}$. In the second shell, we find the symmetry decomposition,
\begin{align}
\label{A1}
\mathbf{\Phi}_{A^{(2)}_{1}}(\mathbf r)&=\frac{1}{\sqrt{3}}\left\{\text{Re} {A}(\mathbf G^{(2)}_1, \mathbf r) + \text{Re} {A}(\mathbf{G}^{(2)}_2, \mathbf r) + \text{Re} {A}(\mathbf{G}^{(2)}_3, \mathbf r)\right\}\\
\label{B1}
\mathbf{\Phi}_{B^{(2)}_2}(\mathbf r)&=\frac{i}{\sqrt{3}}\left\{\text{Im} {A}(\mathbf G^{(2)}_1, \mathbf r) - \text{Im} {A}(\mathbf{G}^{(2)}_2, \mathbf r) + \text{Im} {A}(\mathbf{G}^{(2)}_3, \mathbf r)\right\}
\end{align}
while the two-dimensional irreps are two component vectors
\begin{align}
\label{E1}
\mathbf{\Phi}_{E^{(2)}_1}(\mathbf r)&=\frac{1}{\sqrt{6}}\left\{2\text{Re} {A}(\mathbf G^{(2)}_1, \mathbf r)- \text{Re} {A}(\mathbf{G}^{(2)}_2, \mathbf r)  - \text{Re} {A}(\mathbf{G}^{(2)}_3, \mathbf r),  \sqrt{3}(\text{Re} {A}(\mathbf{G}^{(2)}_2, \mathbf r) - \text{Re} {A}(\mathbf{G}^{(1)}_3, \mathbf r)) \right\}\\
\label{E2}
\mathbf{\Phi}_{E^{(2)}_2}(\mathbf r)&=\frac{i}{\sqrt{6}}\left\{2\text{Im} {A}(\mathbf G^{(2)}_1, \mathbf r)- \text{Im} {A}(\mathbf{G}^{(2)}_2, \mathbf r)  - \text{Im} {A}(\mathbf{G}^{(1)}_3, \mathbf r),  \sqrt{3}(\text{Im} {A}(\mathbf{G}^{(2)}_2, \mathbf r) - \text{Im} {A}(\mathbf{G}^{(1)}_3, \mathbf r)) \right\}
\end{align}
The $A_2$ remains invisible, and requires the third shell of lattice Bragg peaks, but now $B_2$ is visible. If the third shell of lattice Bragg peaks were visible, all irreps would be resolvable including $A_2$, but we do not present results for the third shell as we expect these Bragg peaks are likely beyond experimentally feasible resolution. Instead, in Section \ref{sm:convolution} we will show how to access extinct irreps, including $A_2$, within the first Brillouin shell performing a convolution technique. 

\subsection{Symmetry channels for $\sqrt{3}\times\sqrt{3}$ order}

\begin{figure*}[t]
\centering
\includegraphics[width=\textwidth]{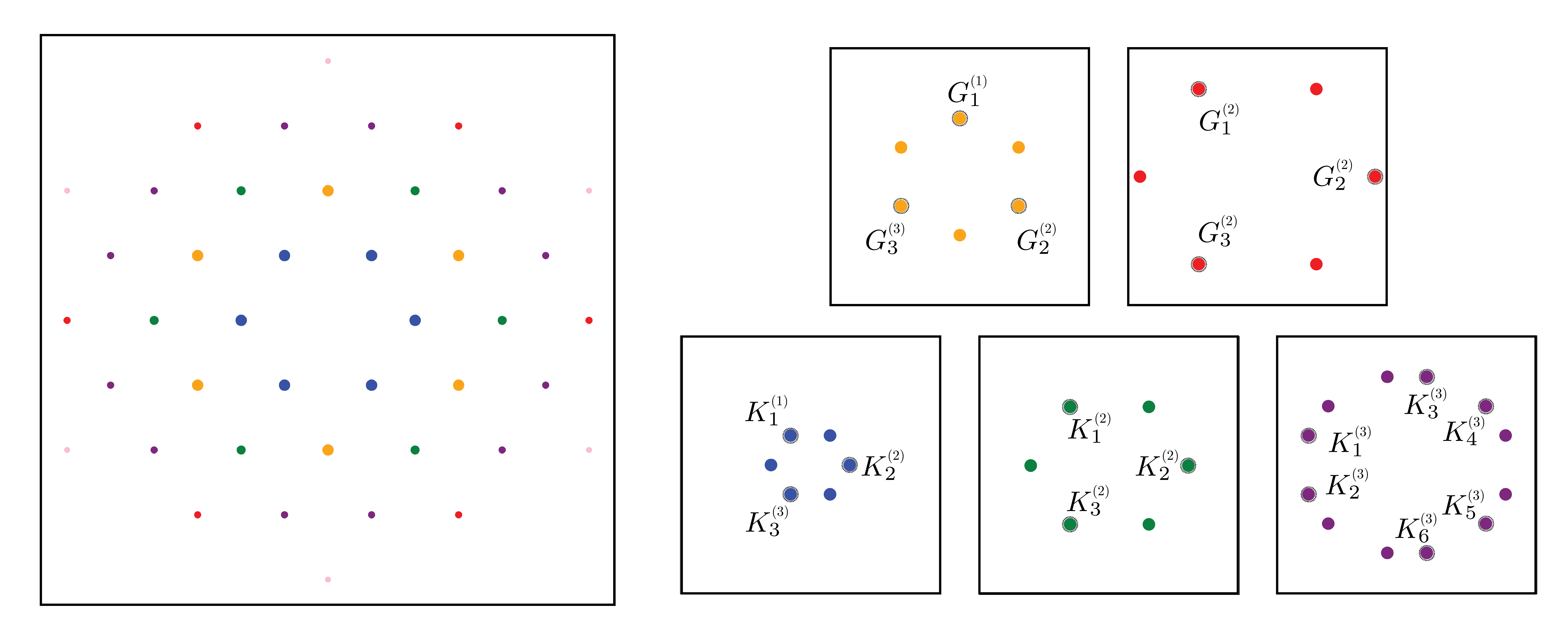}
\vspace{-0.3cm}
\caption{\textbf{Bragg peaks for $\sqrt{3}\times \sqrt{3}$ Kekulé charge density wave.}  Left: Bragg peaks in the first three Brillouin zones for a hexagonal lattice with $\sqrt{3}\times \sqrt{3}$ CDW order, with the dot size decreasing for wavevectors of larger magnitude. Right: definitions of the wavevectors $\mathbf{G}_i^{(n)}$, $\mathbf{K}_i^{(n)}$ colour-coded to match the left image. Orange and red are the first and second shell of lattice Bragg peaks respectively, while blue, green, and purple are the first, second and third shells of CDW Bragg peaks. }
\label{f:supp_fig2}
\vspace{0cm}
\end{figure*}

We now consider the effect of a $\sqrt{3} \times \sqrt{3}$ CDW on a parent system with $C_{6v}$ symmetry. Sometimes referred to as Kekulé order, the $\sqrt{3}\times\sqrt{3}$ order appears in a number of correlated hexagonal systems, including certain kagome metals and multilayer graphene systems \cite{Nuckolls2023, Kim2023, liu2024visualizing, arachchige2022charge, ortiz2024stability, wang2025formation}, and has been proposed to arise from a variety of electron- and phonon-mediated scenarios \cite{ingham2023quadratic, beck2025kekule}. Note that in this case two distinct translations are broken, and so the relevant extended point group is $C^{\prime\prime}_{6v}$, the character table for which is given in the previous section. We show the Bragg peaks in Fig. \ref{f:supp_fig2}, identifying by colour the first three shells of CDW Bragg peaks; the lattice Bragg peaks are unchanged from the previous section.

\subsubsection{First shell}
In the first shell of CDW Bragg peaks, one finds the two- and four-dimensional irreps,
\begin{align}
\label{Ep1}
\mathbf{\Phi}_{E'_1}(\mathbf r)&=\frac{1}{\sqrt{3}}\left\{\text{Re}\left[ {A}(\mathbf K_1, \mathbf r) + {A}(\mathbf K_2, \mathbf r) +  {A}(\mathbf K_3, \mathbf r)\right], \text{Im} \left[ {A}(\mathbf K_1, \mathbf r) +  {A}(\mathbf K_2, \mathbf r) + {A}(\mathbf K_3, \mathbf r)\right] \right\}\\
\label{G}
\mathbf{\Phi}_{G}(\mathbf r)&=\frac{1}{\sqrt{3}}\left\{ \text{Re}\left[ {A}(\mathbf K_1, \mathbf r) + \omega{A}(\mathbf K_2, \mathbf r) +  \omega^*{A}(\mathbf K_3, \mathbf r)\right], \text{Im} \left[ {A}(\mathbf K_1, \mathbf r) +  \omega {A}(\mathbf K_2, \mathbf r) + \omega^* {A}(\mathbf K_3, \mathbf r)\right]  \right. \nonumber \\
& \ \ \ \ \ \  \ \ \ \ \left. \text{Re}\left[ {A}(\mathbf K_1, \mathbf r) + \omega^* {A}(\mathbf K_2, \mathbf r) +  \omega {A}(\mathbf K_3, \mathbf r)\right], \text{Im} \left[ {A}(\mathbf K_1, \mathbf r) +  \omega^*  {A}(\mathbf K_2, \mathbf r) + \omega {A}(\mathbf K_3, \mathbf r)\right]  \right\}
\end{align}
where $\omega = e^{2\pi i/3}$. Ref. \cite{Nuckolls2023} shows three rather than six plots, because there the two components of $E'_1$ are plotted as the real and imaginary parts of a single complex number, represented as a vector field, and the four components of $G$ as two vectors fields. Note though, that Ref. \cite{Nuckolls2023}  groups the $A_1$ and $B_1$ components of the lattice Bragg peaks together as the real and imaginary parts of a complex number -- which our analysis reveals to be a limiting approach, as these two objects are distinct objects according to symmetry, transforming differently under $C_2$, and imaging these irreps separately gives additional symmetry insight.

\subsubsection{Higher-order $\sqrt{3}\times\sqrt{3}$ CDW Bragg peaks}
\label{supp-K-bz3}

The second set of Kekul\'e Bragg peaks are simply twice the first set, while the third shell of CDW Bragg peaks are given by the following explicit expressions: $\{\mathbf{K}^{(3)}_1, \mathbf{K}^{(3)}_2, \mathbf{K}^{(3)}_3, \mathbf{K}^{(3)}_4, \mathbf{K}^{(3)}_5, \mathbf{K}^{(3)}_6 \} = \{\mathbf{G}_1+2\mathbf{G}_2-\mathbf{K}_3, 2\mathbf{G}_1+\mathbf{G}_2+\mathbf{K}_3, \mathbf{G}_1-\mathbf{G}_2-\mathbf{K}_1,-\mathbf{G}_1-2\mathbf{G}_2+\mathbf{K}_3, -2\mathbf{G}_1-\mathbf{G}_2-\mathbf{K}_3, -\mathbf{G}_1+\mathbf{G}_2+\mathbf{K}_2\}$. Since the second shell of CDW Bragg peaks are just twice the first, they unsurprisingly do not really contain any new information from a symmetry perspective.

The new irrep that becomes visible in the third Brillouin zone is the order parameter symmetry $E_2'$ which transforms differently under mirror to the $E_1'$ and $G$ visible in the first Brillouin zone. We simply state the $E_2'$ channel in the third zone rather than give expressions for $E_1'$ and $G$ as well:
\begin{align}
\label{Ep2}
\mathbf{\Phi}_{E'_2}(\mathbf r)&=\frac{1}{\sqrt{3}}\left\{\text{Re}\left[ {A}(\mathbf K^{(3)}_1, \mathbf r) + {A}(\mathbf K^{(3)}_2, \mathbf r) +  {A}(\mathbf K^{(3)}_3, \mathbf r)+{A}(\mathbf K^{(3)}_4, \mathbf r) + {A}(\mathbf K^{(3)}_5, \mathbf r) +  {A}(\mathbf K^{(3)}_6, \mathbf r)\right], \right. \nonumber \\
& \ \ \ \ \ \  \ \ \ \ \left.\text{Im} \left[ {A}(\mathbf K^{(3)}_1, \mathbf r) + {A}(\mathbf K^{(3)}_2, \mathbf r) +  {A}(\mathbf K^{(3)}_3, \mathbf r)+{A}(\mathbf K^{(3)}_4, \mathbf r) + {A}(\mathbf K^{(3)}_5, \mathbf r) +  {A}(\mathbf K^{(3)}_6, \mathbf r) \right] \right\}
\end{align}

\newpage

\section{Symmetry convolution technique}
\label{supp-convolution}

As explained in the previous section, certain symmetry breaking patterns are encoded strictly in Bragg peaks of higher Brillouin zones. Here we describe a procedure by which this information can be encoded in the first Brillouin zone of manipulated data. We refer to this as the ``symmetry convolution technique.''

The idea is to convolve the STM data with the basis function of a particular irrep, $\varphi_\Gamma(\mathbf r)$ -- which we refer to as a ``symmetry-mask'' -- producing new Fourier transformed data $A'(\mathbf M_i, \mathbf r)$,
\begin{align}
\label{conv}
A'(\mathbf M_i, \mathbf r) = \int d^2 \mathbf{r}^{\prime} e^{-i \mathbf M_i\cdot(\mathbf r^{\prime}-\mathbf r)} \varphi_\Gamma(\mathbf r') s\left(\mathbf r^{\prime}\right) w\left(\mathbf r^{\prime}-\mathbf r\right),
\end{align}
The first Brillouin zone of the symmetry-convolved data will contain information related to $\chi(\mathbf r)$, which otherwise may have been extinct.

As a simple example, consider a CDW pattern in an $F_2$ irrep, which is odd under the mirror symmetries $\sigma_v, \sigma_d$. Suppose an STM signal contains a lattice contribution $s_0(\mathbf r)$, i.e. without symmetry breaking, as well as an $F_2$ contribution, with some relative magnitudes $\alpha$ and $\beta$
\begin{align}
  s(\mathbf r)&=\alpha s_0(\mathbf r) + \beta s_{F_2}(\mathbf r).
 \end{align}
 Using an $ \varphi_{A_2}(\mathbf r)$ basis function, we may represent the $s_{F_2}(\mathbf r)$ as
 \begin{align}
s_{F_2}(\mathbf r)= \varphi_{A_2}(\mathbf r) \sum_{i=1}^3\cos(\mathbf M_i\cdot\mathbf r).
 \end{align}
As discussed, Fourier transforming and applying the symmetry decomposition, this $F_2$ contribution cannot be imaged from the first Brillouin zone peaks. However, considering the local Fourier amplitude of the modified data Eq. \eqref{conv}
\begin{align}
    s'(\mathbf r)= \chi_{A_2}(\mathbf r) s(\mathbf r)=\alpha \chi_{A_2}(\mathbf r) s_0(\mathbf r) + \beta[\chi_{A_2}(\mathbf r) ]^2\sum_{i=1}^3\cos(\mathbf M_i\cdot\mathbf r)
\end{align}
the term proportional to $\beta$ now transforms as $F_1$, allowing it to be imaged in the first Brillouin zone. On the other hand, the first term is odd under mirror symmetry and is now extinct in the first Brillouin zone.

More generally, of the CDW orders $F_1,F_2,F_3,F_4$, only $F_1$ and $F_3$ are visible in the first $M$-point shell as we have previously derived
\begin{align}
\notag \mathbf{\Phi}_{F_{1}}(\mathbf r)&=\text{Re}\left\{ {A}(\mathbf M_1, \mathbf r), {A}(\mathbf M_2, \mathbf r),  {A}(\mathbf M_3, \mathbf r)\right\}\\
\notag \mathbf{\Phi}_{F_{2}}(\mathbf r)&=0\\
\notag \mathbf{\Phi}_{F_{3}}(\mathbf r)&=\text{Im}\left\{ {A}(\mathbf M_1, \mathbf r), {A}(\mathbf M_2, \mathbf r),  {A}(\mathbf M_3, \mathbf r)\right\}\\
\mathbf{\Phi}_{F_{4}}(\mathbf r)&=0.
\end{align}
However, using the multiplication rules, one finds that convolution with either $B_1$ or $B_2$ allows for $F_2$ and $F_4$ to be resolved via the convolution approach, since
\begin{align}
B_1\otimes \{F_1,F_2,F_3,F_4\}&=\{F_3,F_4,F_1,F_2\},\\
B_2\otimes \{F_1,F_2,F_3,F_4\}&=\{F_4,F_3,F_2,F_1\},
\end{align}
then taking the convolution of either of $\chi_{B_1}(\mathbf r)$ or $\chi_{B_2}(\mathbf r)$ with the STM function $s(\mathbf r)$ will allow for $F_2,F_4$ to be probed. 

To implement these convolutions, we shall explicitly define symmetry masks for $B_1$ and $B_2$ irreps, though our concept is applicable more broadly. First, we define the lattice periodic functions
\begin{align}
X(\mathbf r)&=\frac{2}{3} \left(\cos (2 \pi  x) \sin \left(\tfrac{2 \pi 
   y}{\sqrt{3}}\right)+\sin \left(\tfrac{4 \pi 
   y}{\sqrt{3}}\right)\right),\quad
Y(\mathbf r)=\frac{2}{\sqrt{3}} \sin (2 \pi  x) \cos \left(\tfrac{2 \pi 
   y}{\sqrt{3}}\right).
\end{align}
and the symmetry masks we shall use for convolution,
\begin{align}
\varphi_{B_1}(\mathbf r)&=X(\mathbf r)\left[X(\mathbf r)^2-3Y(\mathbf r)^2\right],\\
\varphi_{B_2}(\mathbf r)&=Y(\mathbf r)\left[3X(\mathbf r)^2-Y(\mathbf r)^2\right]
\end{align}
Finally, we add a normalization factor such that
\begin{align}
\int_{UC} d^2r\left[\chi_{\Gamma}(\mathbf r)\right]^2=1.
\end{align}
where integration is restricted to the unit cell, $UC$.

\begin{figure}[t!]
\centering
\includegraphics[width=0.35\textwidth]{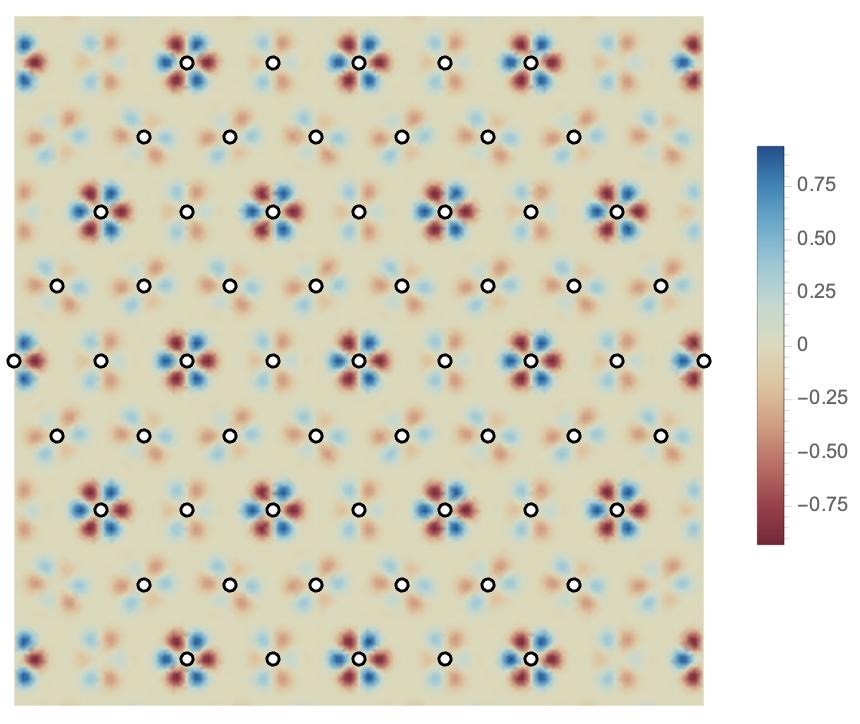}\hspace{1cm}
\includegraphics[width=0.35\textwidth]{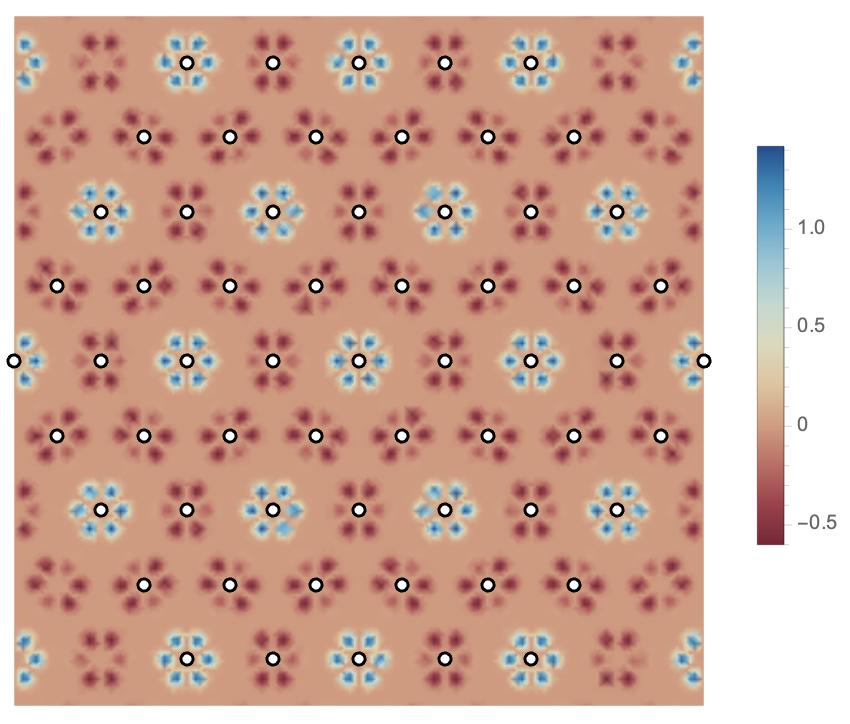}
\caption{ \textbf{Symmetry convolution method.} Left: Density map of a function basis function $f(\mathbf r)\in F_4$. White circles are the original kagome lattice sites. Right: Density map of a function $\chi_{B_2}(\mathbf r)\times f(\mathbf r)$; the resulting image transforms trivially under all reflections and rotations, and hence belongs to $F_1$. Hence, imaging $F_1$ on this convolved data allows us to detect the presence of $F_2$ in the original data.}
    \label{fig:mask}
\end{figure}

We illustrate the symmetry convolution procedure explicitly in Fig. \ref{fig:mask}; on the left we show a real space plot of function $f(
\mathbf r)$ which transforms as $F_4$, and on the right we show that same function multiplied by $\varphi_{B_2}(\mathbf r)$, which is clearly seen to transform as $F_1$. Hence if STM data $s(\mathbf r)$ contains $F_4$, then the local Fourier amplitude of $\varphi_{B_2}(\mathbf r) s(\mathbf r)$ would return a non-zero $\mathbf \Phi_{F_1}(\mathbf r)$;  the appearance of $\mathbf \Phi_{F_1}(\mathbf r)$ in $s'(\mathbf r)$ directly implies the existence of $\mathbf \Phi_{F_4}(\mathbf r)$ in $s(\mathbf r)$.

\newpage
\section{Window function}\label{supp-window}

\begin{figure}[t!]
\centering
\includegraphics[width=0.95\textwidth]{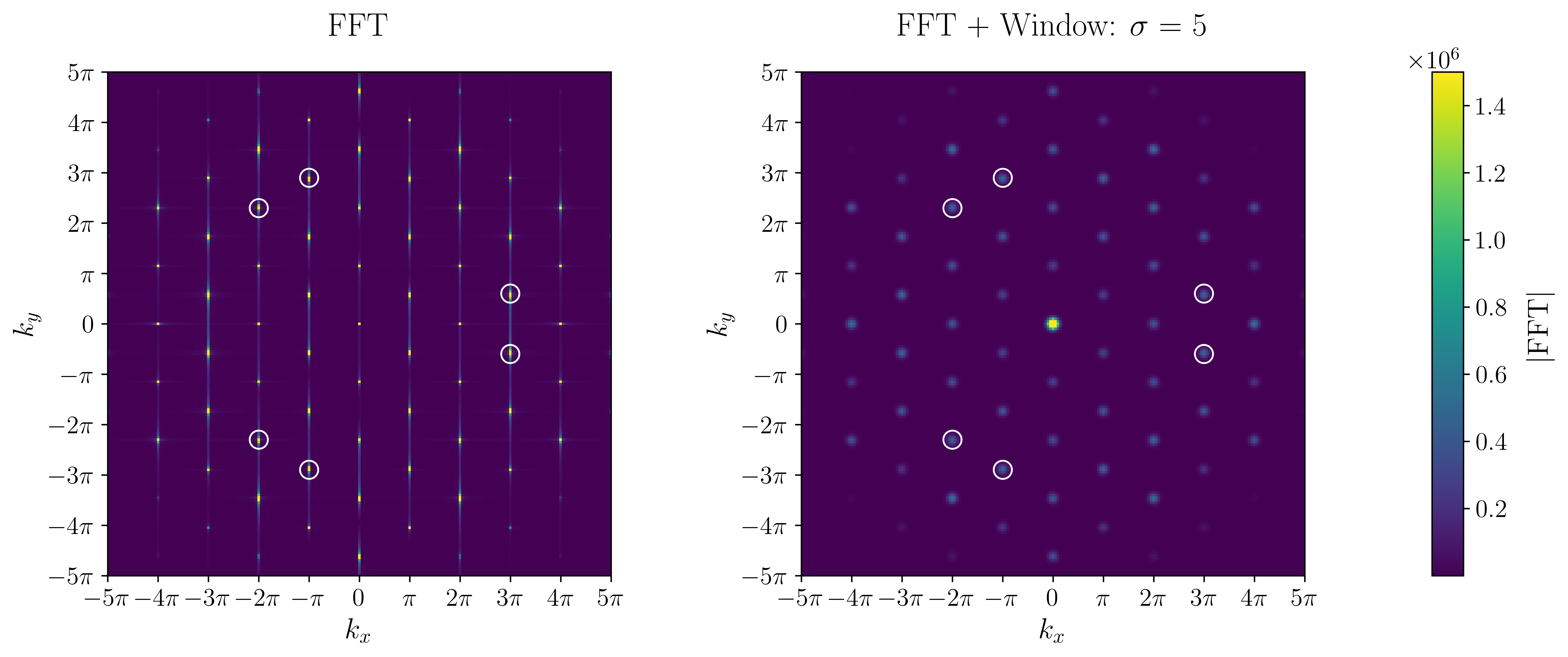}
\caption{\textbf{Effect of the windowing function on Fourier transformed STM data.} FFT of $F_4$ flux order without (left) and with (right) real-space convolution with a Gaussian window function of width $\sigma$.}
    \label{fig:FFT_window}
\end{figure}

Here we discuss the use of window functions to process STM data to reduce errors from discrete sampling, preventing ``spectral leakage'' into spurious symmetry breaking channels.

In STM experiments, the measured field of view is typically defined over a square domain, leading to effectively sharp boundaries in real space. These abrupt edges act as a rectangular window function, which introduces artificial discontinuities at the boundary. In Fourier space, this manifests as spectral leakage: the Fourier peaks acquire extended sinc-like tails that decay slowly and anisotropically. Crucially, this leakage exhibits a strong directional bias -- the tails preferentially extend along the axes aligned with the real-space window (typically the $x$ and $y$ directions). As a result, the intrinsic $C_6$ symmetry of the underlying physics becomes obscured by a spurious nematicity imposed by the sampling geometry. To mitigate this, we apply a smooth Gaussian envelope to the real-space data -- as in Eq. \eqref{main:stm} -- effectively softening the boundary. This transforms the sharp-window Fourier transform into a Gaussian-convolved spectrum, yielding compact, isotropic Fourier peaks. The resulting suppression of anisotropic leakage approximately restores the expected hexagonal structure in momentum space and improves the fidelity of symmetry-based decompositions.

To explore the precision with which symmetry channels can be decomposed, we first generate a CDW pattern that is purely in the symmetry $F_1$ channel, and take the FFT. We find that, due to sampling, the peaks are unequal in magnitude, which appears as leakage into $F_2$ (the other CDW with purely real peaks).  We address this using a Gaussian window function of width $\sigma$, 
\begin{align}
\label{window_func}
w(\mathbf r) = \exp\left(-\frac{r^2}{2\sigma^2}\right),
\end{align}
to smooth out the sampling.
See Fig. \ref{fig:FFT_window} for a demonstration.

\clearpage 

\section{Real space decomposition}
\label{supp-real_space}

\begin{figure}[t]
    \centering
\includegraphics[width=0.2\linewidth]{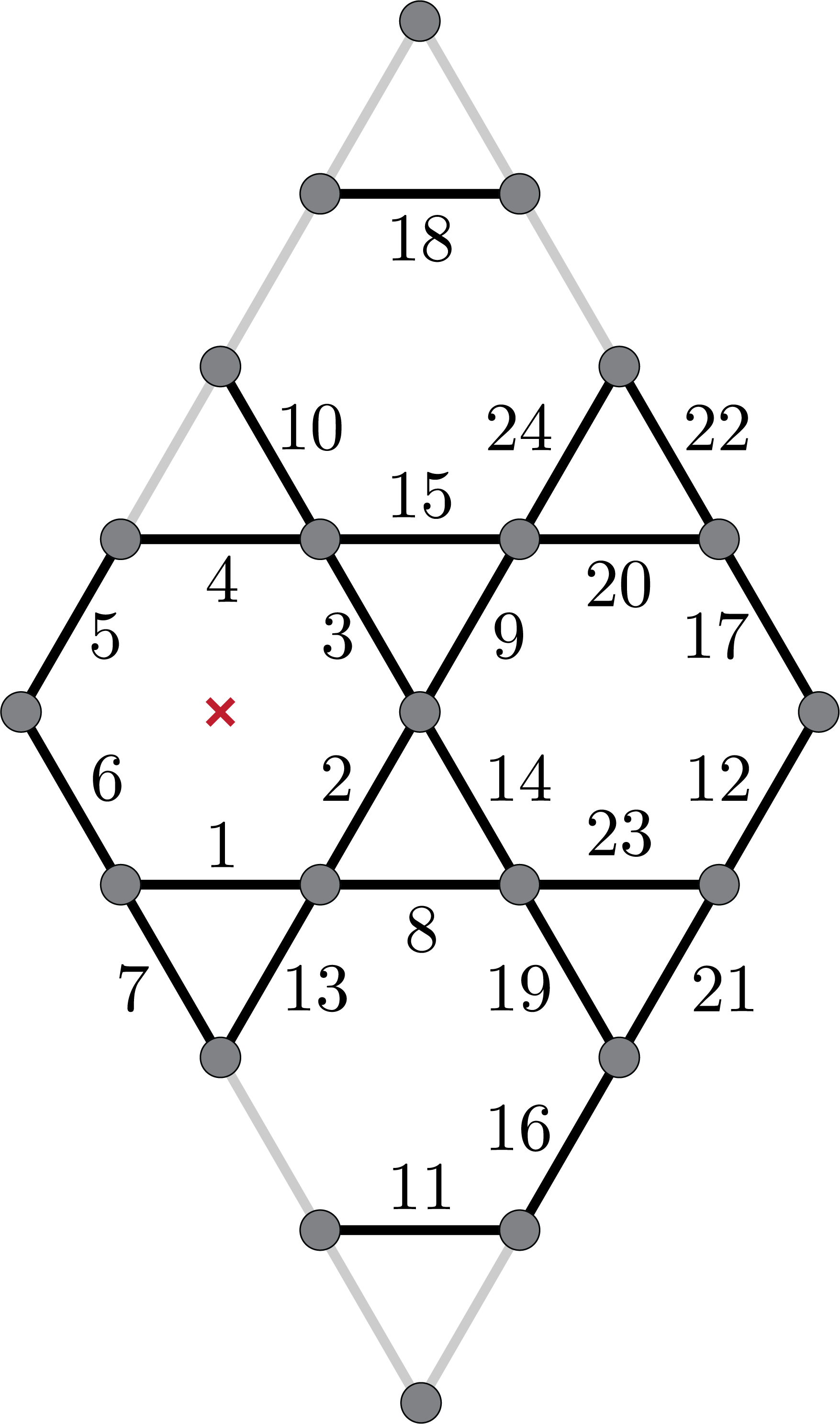}
    \caption{\textbf{Real space basis defined on via bonds of the CDW unit cell.} Dark lines mark the 24 unique bonds in the $2\times2$ unit cell, indexed by a number from 1 to 24. These points form a 24 dimensional reducible representation, which can be decomposed using the group projector formalism. The red cross marks the origin. }
    \label{fig:midpoints}
\end{figure}

The approach we have adopted is to take a subset of the STM data as a basis and then use the equations which describe how these basis elements transform into each other to decompose them into symmetry-distinct combinations. While we chose this subset to comprise the Bragg peaks of the lattice and CDW, this basic idea does not require that we work with momentum space data. In this section, we explore an alternative approach which works purely in real space. The idea is to evaluate the STM data on the bonds in the enlarged unit cell, and use this as a basis.

In practice, one can analyse STM data by scanning the system and identifying the center of the CDW unit cell everywhere on the surface, then performing the symmetry decomposition in each individual unit cell using the approach we set out in this section. This method seems more natural in cases where there is strong nanoscale texturing of the order parameter. Naturally, the method depends on the underlying lattice; in keeping with our earlier results, we present an explicit demonstration of the technique for the example of the kagome lattice with $2\times 2$ ordering.

\subsection{Real space bond decomposition for the kagome lattice with $2\times 2$ order}

To decompose the STM image into irreps, we evaluate the inner product of the projection operator eigenvectors with a 24-dimensional STM vector: $\mathbf S$, which is the STM map evaluated at the 24 unique bonds as shown in Fig. \ref{fig:midpoints}. We denote this scalar product as the {\it weight} ${\cal W}_n^\Gamma$,
\begin{align}
{\cal W}_n^\Gamma = {\cal P}^\Gamma_n\cdot \mathbf S.
\end{align}
Denoting the number of non-zero, real eigenvalues as $N_\Gamma$, then
for 1D irreps, the number of $N$ corresponds to the multiplicity. For 2D/3D irreps, $N$ counts the components; the multiplicity for e.g. a 3D irrep is then $N/3$. We find the explicit expressions
\begin{align}
    {\cal P}_1^{A_1} & = (1,1,1,1,1,1,1,1,1,1,1,1,1,1,1,1,1,1,1,1,1,1,1,1)\\
    {\cal P}_1^{B_1} & = (-1,1,-1,1,-1,1,-1,1,-1,1,-1,1,-1,1,-1,1,-1,1,-1,1,-1,1,-
   1,1) \\
    \begin{bmatrix}{\cal P}_1^{E_1}\\{\cal P}_2^{E_1} \end{bmatrix} & = \begin{bmatrix}(0,1,1,0,-1,-1,1,0,-1,-1,0,1,-1,-1,0,1,1,0,1,0,-1,-1,0,1)\\(1,0,-1,-1,0,1,-1,-1,0,1,1,0,0,1,1,0,-1,-1,-1,-1,0,1,1,0) \end{bmatrix}\\
\begin{bmatrix}{\cal P}_1^{E_2}\\{\cal P}_2^{E_2} \end{bmatrix} & = \begin{bmatrix}(0,1,-1,0,1,-1,-1,0,1,-1,0,1,1,-1,0,1,-1,0,-1,0,1,-1,0,1)\\(1,0,-1,1,0,-1,-1,1,0,-1,1,0,0,-1,1,0,-1,1,-1,1,0,-1,1,0)\end{bmatrix}\\
\begin{bmatrix}{\cal P}_{\mathbf M_1}^{F_{1a}}\\{\cal P}_{\mathbf M_2}^{F_{1a}} \\ {\cal P}_{\mathbf M_3}^{F_{1a}} \end{bmatrix} & = \begin{bmatrix}({-1, -1, -1, -1, -1, -1, 1, 0, 0, 1, 0, 0, 0, 0, 1, 0, 0, 1, 0, 0, 1, \
0, 0, 1})\\({-1, -1, -1, -1, -1, -1, 0, 0, 1, 0, 0, 1, 0, 1, 0, 0, 1, 0, 0, 1, 0, \
0, 1, 0})\\ ({-1, -1, -1, -1, -1, -1, 0, 1, 0, 0, 1, 0, 1, 0, 0, 1, 0, 0, 1, 0, 0, \
1, 0, 0}) \end{bmatrix}\\
\begin{bmatrix}{\cal P}_{\mathbf M_1}^{F_{1b}}\\{\cal P}_{\mathbf M_2}^{F_{1b}} \\ {\cal P}_{\mathbf M_3}^{F_{1c}} \end{bmatrix} & = \frac{\sqrt{3}}{\sqrt{2}}\begin{bmatrix}({0, 0, 0, 0, 0, 0, -1, 0, 0, -1, 0, 0, 0, 0, -1, 0, 0, -1, 1, 1, 0, \
1, 1, 0})\\({0, 0, 0, 0, 0, 0, 0, 0, -1, 0, 0, -1, 0, -1, 0, 0, -1, 0, 1, 0, 1, \
1, 0, 1})\\ ({0, 0, 0, 0, 0, 0, 0, -1, 0, 0, -1, 0, -1, 0, 0, -1, 0, 0, 0, 1, 1, \
0, 1, 1}) \end{bmatrix}\\
\begin{bmatrix}{\cal P}_{\mathbf M_1}^{F_{2}}\\{\cal P}_{\mathbf M_2}^{F_{2}} \\ {\cal P}_{\mathbf M_3}^{F_{2}} \end{bmatrix} & = \frac{\sqrt{3}}{2}\begin{bmatrix}({1, 0, -1, 1, 0, -1, -1, -1, 0, -1, -1, 0, 0, 1, 1, 0, 1, 1, 1, -1, \
0, 1, -1, 0})\\({0, -1, 1, 0, -1, 1, -1, 0, -1, -1, 0, -1, 1, 1, 0, 1, 1, 0, -1, 0, \
1, -1, 0, 1})\\ ({-1, 1, 0, -1, 1, 0, 0, -1, -1, 0, -1, -1, 1, 0, 1, 1, 0, 1, 0, 1, \
-1, 0, 1, -1}) \end{bmatrix}\\
\begin{bmatrix}{\cal P}_{\mathbf M_1}^{F_{3a}}\\{\cal P}_{\mathbf M_2}^{F_{3a}} \\ {\cal P}_{\mathbf M_3}^{F_{3a}} \end{bmatrix} & = \begin{bmatrix}({-1, 1, -1, 1, -1, 1, 1, 0, 0, -1, 0, 0, 0, 0, 1, 0, 0, -1, 0, 0, 1, \
0, 0, -1})\\({-1, 1, -1, 1, -1, 1, 0, 0, 1, 0, 0, -1, 0, -1, 0, 0, 1, 0, 0, -1, 0, \
0, 1, 0})\\ ({-1, 1, -1, 1, -1, 1, 0, -1, 0, 0, 1, 0, 1, 0, 0, -1, 0, 0, 1, 0, 0, \
-1, 0, 0}) \end{bmatrix}\\
\begin{bmatrix}{\cal P}_{\mathbf M_1}^{F_{3b}}\\{\cal P}_{\mathbf M_2}^{F_{3b}} \\ {\cal P}_{\mathbf M_3}^{F_{3b}} \end{bmatrix} & = \frac{\sqrt{3}}{2\sqrt{2}}\begin{bmatrix}({2, -2, 0, -2, 2, 0, 1, 0, -2, -1, 0, 2, 0, 0, -1, 0, 0, 1, -1, 1, 0, \
1, -1, 0})\\({2, 0, 2, -2, 0, -2, 0, 2, 1, 0, -2, -1, 0, 1, 0, 0, -1, 0, -1, 0, \
-1, 1, 0, 1})\\ ({0, -2, 2, 0, 2, -2, -2, -1, 0, 2, 1, 0, -1, 0, 0, 1, 0, 0, 0, 1, -1, \
0, -1, 1}) \end{bmatrix}\\
\begin{bmatrix}{\cal P}_{\mathbf M_1}^{F_{4}}\\{\cal P}_{\mathbf M_2}^{F_{4}} \\ {\cal P}_{\mathbf M_3}^{F_{4}} \end{bmatrix} & = \frac{1}{\sqrt{2}}\begin{bmatrix}({-1, 0, 1, 1, 0, -1, 1, -1, 0, -1, 1, 0, 0, 1, -1, 0, -1, 1, -1, -1, \
0, 1, 1, 0})\\({0, -1, -1, 0, 1, 1, 1, 0, 1, -1, 0, -1, -1, 1, 0, 1, -1, 0, 1, 0, \
-1, -1, 0, 1})\\ ({1, 1, 0, -1, -1, 0, 0, -1, 1, 0, 1, -1, -1, 0, -1, 1, 0, 1, 0, 1, 1, \
0, -1, -1}) \end{bmatrix}
\end{align}
We have used square braces $[ \cdot ]$ to group the components of 2D and 3D irreps. As an example, to evaluate the component of $F_2$ in the CDW unit cell in terms of the STM data evaluated at bond $i$, denoted $S_i$, the above formulae yield
\begin{align}
    \begin{pmatrix} \mathcal{W}_{\mathbf{M}_1} \\ \mathcal{W}_{\mathbf{M}_2} \\ \mathcal{W}_{\mathbf{M}_3} \end{pmatrix}_{F_2} &=
    \frac{\sqrt{3}}{2}\begin{pmatrix}
        S_1 - S_3 + S_4 - S_6 - S_7 - S_8 - S_{10} - S_{11} + S_{14} + S_{15} + S_{17} + S_{18} + S_{19} - S_{20} + S_{22} - S_{23} \\
        -S_2 + S_3 - S_5 + S_6 - S_7 - S_9 - S_{10} - S_{12} + S_{13} + S_{14} + S_{16} + S_{17} - S_{19} + S_{21} - S_{22} + S_{24} \\
        -S_1 + S_2 - S_4 + S_5 - S_8 - S_9 - S_{11} - S_{12} + S_{13} + S_{15} + S_{16} + S_{18} + S_{20} - S_{21} + S_{23} - S_{24}
    \end{pmatrix}
\end{align}

\newpage

\section{Application to synthetic STM data}
\label{supp-synth-stm}

\subsection{Calculating STM data from tight-binding models}

The differential tunneling current imaged in STM is proportional to LDOS$(\text{eV},\mathbf{r})$, where eV is the bias voltage, and the local density of states (LDOS) is given by
\begin{align}
    \text{LDOS}(E,\mathbf{r}) = - \tfrac{1}{\pi} \text{Im} \langle c_\mathbf{r}c^\dag_\mathbf{r}\rangle_E = -\tfrac{1}{\pi} \text{Im} G(E, \mathbf{r})
\end{align}
where $c^\dag_\mathbf{r}$ is the electron creation operator and $G(E,\mathbf{r})$ is the electron Green' s function.  To relate the continuum electronic density of states to the wavefunctions of a lattice model, we expand the continuum electron operator in terms of lattice creation operators associated to an orbital $\sigma$ and a lattice site  $\mathbf{R}$, via $ c^\dag_{\mathbf{r}} = \sum_{\sigma\mathbf{R}} w_{\sigma\mathbf{R}}(\mathbf{r}) c^\dag_{\sigma\mathbf{R}}$ where $w_{\sigma\mathbf{R}}(\mathbf{r})$ are a set of real space orbitals indexed by $\sigma$ and centered at the lattice sites $\mathbf{R}$, e.g. \cite{kreisel2021quasi, sobral2023machine, rhodes2024probing, nag2024pomeranchuk, holbrook2024real}. One obtains
\begin{align}
    \text{LDOS}(E,\mathbf{r}) &= - \tfrac{1}{\pi} \text{Im} \sum_{\sigma\sigma',\mathbf{R}\mathbf{R}'} w^*_{\sigma\mathbf{R}}(\mathbf{r}) w_{\sigma'\mathbf{R}'}(\mathbf{r}) \langle c_{\sigma\mathbf{R}} c^\dag_{\sigma'\mathbf{R}'}\rangle_E \nonumber\\
  &= - \tfrac{1}{\pi} \text{Im} \sum_{\sigma\sigma',\mathbf{R}\mathbf{R}'} w^*_{\sigma\mathbf{R}}(\mathbf{r}) w_{\sigma'\mathbf{R}'}(\mathbf{r}) G(E, \mathbf{R}-\mathbf{R}')_{\sigma\sigma'}
\end{align}
where $G(E, \mathbf{R}-\mathbf{R}')_{\sigma\sigma'}$ is the Green's function of the lattice model,
\begin{gather}
\label{tb_green}
G(E, \mathbf{R}-\mathbf{R}')_{\sigma\sigma'} = \int \frac{d^d\mathbf{k}}{(2\pi)^d} \left(\frac{1}{E - \mathcal{H}(\mathbf{k}) + i0} \right)_{\sigma\sigma'} e^{i\mathbf{k}(\mathbf{R}-\mathbf{R}')}
\end{gather}
where $\mathcal{H}(\mathbf{k})_{\sigma\sigma'}$ is the tight-binding Hamiltonian, defined as a matrix acting in orbital basis. In all calculations to follow, we shall employ Gaussian model $s$-orbitals for $w_{\sigma\mathbf{R}}(\mathbf{r})$.

\subsection{Synthetic STM in the presence of spatial modulation}

To model the influence of the charge density wave, we construct a model which takes a set of momentum transfers we term k-points, $\mathbf{Q}_i$, and band-folds the dispersion by working with a kinetic Hamiltonian $\text{diag}(\mathcal{H}(\mathbf{k}), \mathcal{H}(\mathbf{k}+\mathbf{Q}_1), ...)$. Then we add a tunneling matrix which is off-diagonal in k-point space, hybridising the folded bands. The two k-points $\mathbf{Q}_i$ and $\mathbf{Q}_j$ are connected by a CDW matrix element $T(\mathbf{Q}_i-\mathbf{Q}_j)_{\sigma_1\sigma_2}$ which can be a matrix in sublattice space. 

The continuum electron operator is then also written in terms of k-points indexed by $\alpha$,
\begin{align}
   c^\dag_{\mathbf{r}} = \sum_{\mathbf{R}} w_{\sigma\mathbf{R}}(\mathbf{r})c^\dag_{\sigma \mathbf{R}} = \sum_{\mathbf{R},\mathbf{k},\alpha} w_{\sigma\mathbf{R}}(\mathbf{r}) e^{i(\mathbf{k}+\mathbf{Q}_\alpha)\cdot \mathbf{R}} c^\dag_{\sigma \mathbf{k},\alpha} 
\end{align}
where $\alpha$ indexes k-points, and $\mathbf{k}$ denotes momentum modulo the reciprocal lattice vectors of the expanded unit cell, i.e. quasimomentum within the resultant mini Brillouin zone. Fourier transforming as above, one finds
\begin{align}
\text{LDOS}(E, \mathbf{r})  = -\tfrac{1}{\pi}\text{Im}\sum_{\mathbf{R}\mathbf{R}',\mathbf{k}\mathbf{k}',\alpha\alpha',\sigma\sigma'} w_{\sigma\mathbf{R}}(\mathbf{r}) w_{\sigma'\mathbf{R}'}(\mathbf{r}) \,e^{i(\mathbf{k}\cdot \mathbf{R}-\mathbf{k}'\cdot \mathbf{R}')} e^{i(\mathbf{Q}_\alpha\cdot \mathbf{R}-\mathbf{Q}_{\alpha'}\cdot \mathbf{R}')}\,\langle c_{\sigma\mathbf{k}\alpha}c^\dag_{\sigma'\mathbf{k}'\alpha'}\rangle
\end{align}
Modified translational invariance implies the Green's function is diagonal in folded quasimomentum $\mathbf{k}$, implying 
\begin{align}
   \text{LDOS}(E, \mathbf{r}) = -\tfrac{1}{\pi}\text{Im}\sum_{\mathbf{R}\mathbf{R}',\sigma\sigma'} w_{\sigma\mathbf{R}}(\mathbf{r}) w_{\sigma'\mathbf{R}'}(\mathbf{r})\,\mathcal{G}^{\text{CDW}}_{\sigma\sigma'}(E;\mathbf{R},\mathbf{R}') 
\end{align}
where the CDW Green's function is 
\begin{align}
\label{cdw_green}
\mathcal{G}^{\text{CDW}}_{\sigma\sigma'}(E;\mathbf{R},\mathbf{R}') \equiv \sum_{\mathbf{k},\alpha\alpha'} e^{i\mathbf{k}\cdot (\mathbf{R}-\mathbf{R}')} e^{i(\mathbf{Q}_\alpha\cdot \mathbf{R}-\mathbf{Q}_{\alpha'}\cdot \mathbf{R}')}\,\langle c_{\sigma\mathbf{k}\alpha}c^\dag_{\sigma'\mathbf{k}\alpha'}\rangle
\end{align}

The CDW Green's function is computed by inverting the mean field Hamiltonian in sublattice and k-point space and summing over indices appropriately. We note the similarity of this calculation to STM simulations of moir\'e graphene \cite{Calugaru2022, Hong2022}, in which the moir\'e potential plays the role of the spatial modulation.

\subsection{Subtleties due to gauge choices}

In Eq. \eqref{tb_green}, we work with a gauge in which the electron operator with momentum $\mathbf{k}$ at sublattice $\sigma$ is related to the real space electron operator via
\begin{align}
   c^\dag_{\sigma \mathbf{R}} = \sum_{\mathbf{R},\mathbf{k}} e^{i\mathbf{k}\cdot \mathbf{R}} c^\dag_{\sigma \mathbf{k}} 
\end{align}
In this gauge, the nearest-neighbour tight-binding Hamiltonian reads
\begin{align}
    \mathcal{H}_0(\mathbf{k}) =  -t\begin{pmatrix}
0 & 1+ e^{-i\mathbf{k}\cdot\mathbf{R}_2} & e^{i\mathbf{k}\cdot\mathbf{R}_1}+ e^{-i\mathbf{k}\cdot\mathbf{R}_2} \\
 1+ e^{i\mathbf{k}\cdot\mathbf{R}_2}& 0 & 1+ e^{i\mathbf{k}\cdot\mathbf{R}_1} \\
e^{-i\mathbf{k}\cdot\mathbf{R}_1}+ e^{i\mathbf{k}\cdot\mathbf{R}_2} & 1+ e^{-i\mathbf{k}\cdot\mathbf{R}_1} & 0 \\
\end{pmatrix}
\end{align}
We compare this with the gauge in which the momentum space electron operator is given by
\begin{align}
   c^\dag_{\sigma \mathbf{R}} = \sum_{\mathbf{R},\mathbf{k}} e^{i\mathbf{k}\cdot( \mathbf{R}+\mathbf{r}_\sigma)} c^\dag_{\sigma \mathbf{k}} 
\end{align}
the difference being the sublattice dependent phase $e^{i\mathbf{k}\cdot\mathbf{r}_\sigma}$ has been absorbed into the electron operator via a gauge transformation. In this basis, the nearest-neighbour tight-binding Hamiltonian reads
\begin{align}
    \mathcal{H}_0(\mathbf{k}) =  -t\begin{pmatrix}
0 & e^{i\mathbf{k}\cdot(\mathbf{r}_b-\mathbf{r}_a)}+ e^{i\mathbf{k}\cdot(\mathbf{r}_b-\mathbf{r}_a-\mathbf{R}_2)} & e^{i\mathbf{k}\cdot(\mathbf{r}_b-\mathbf{r}_a+\mathbf{R}_1)}+ e^{i\mathbf{k}\cdot(\mathbf{r}_b-\mathbf{r}_a-\mathbf{R}_2}) \\
 e^{i\mathbf{k}\cdot(\mathbf{r}_a-\mathbf{r}_b)}+ e^{i\mathbf{k}\cdot(\mathbf{r}_a-\mathbf{r}_b+\mathbf{R}_2)}& 0 & e^{i\mathbf{k}\cdot(\mathbf{r}_b-\mathbf{r}_a)}+ e^{i\mathbf{k}\cdot\mathbf{R}_1} \\
e^{i\mathbf{k}\cdot(\mathbf{r}_a-\mathbf{r}_b-\mathbf{R}_1)}+ e^{i\mathbf{k}\cdot\mathbf{r}_a-\mathbf{r}_b+\mathbf{R}_2)} & e^{i\mathbf{k}\cdot(\mathbf{r}_a-\mathbf{r}_b)}+ e^{i\mathbf{k}\cdot(\mathbf{r}_a-\mathbf{r}_b+\mathbf{R}_1)} & 0 \\
\end{pmatrix}
\end{align}

We choose the first of the two gauges in Eq. \eqref{cdw_green} -- note that when working with a CDW order parameter, the Hamiltonian possesses the k-point quantum number $\alpha$ as well; removing the sublattice phases means multiplying the electron operator by $e^{i(\mathbf{k}+\mathbf{Q}_\alpha)\cdot \mathbf{r}_\sigma }$. Therefore, a spatial modulation in the second gauge is related to the first via
\begin{align}
    T(\mathbf{Q}_i-\mathbf{Q}_j)_{\sigma\sigma'}\rightarrow e^{i \mathbf{k}\cdot(\mathbf{r}_\sigma-\mathbf{r}_{\sigma'})} e^{i\mathbf{Q}_i\cdot \mathbf{r}_\sigma-i\mathbf{Q}_j\cdot \mathbf{r}_{\sigma'}} T(\mathbf{Q}_i-\mathbf{Q}_j)_{\sigma\sigma'}
\end{align}
We caution that one must take care to choose a consistent gauge throughout all steps of an STM simulation; failing to do so produces spurious symmetry breaking.

\subsection{Minimal models of $2\times 2$ charge order}

In this section we give explicit details of the tight-binding Hamiltonians used to compute synthetic STM data for a kagome lattice with translationally invariant mirror-breaking orders, as well as $2\times 2$ CDWs. The Hamiltonian consists of a symmetric nearest-neighbour tight-binding model perturbed by an order parameter of a given symmetry via the procedure outlined in the previous section.

\subsubsection{Tight-binding Hamiltonian}
For the symmetric parent Hamiltonian, we take the simple nearest-neighbour tight-binding model of the kagome lattice, defined in the gauge
\begin{align}
{\cal H}_0 = -2t 
\begin{pmatrix}
0 & \cos(\mathbf{k} \cdot \mathbf{d}_{AB}) & \cos(\mathbf{k} \cdot \mathbf{d}_{AC}) \\
\cos(\mathbf{k} \cdot \mathbf{d}_{AB}) & 0 & \cos(\mathbf{k} \cdot \mathbf{d}_{BC}) \\
\cos(\mathbf{k} \cdot \mathbf{d}_{AC}) & \cos(\mathbf{k} \cdot \mathbf{d}_{BC}) & 0
\end{pmatrix}.
\end{align}

\subsubsection{Bond order charge density waves}\label{sec:bondorders}

\begin{figure*}[t]
\centering
\includegraphics[width=0.95\textwidth]{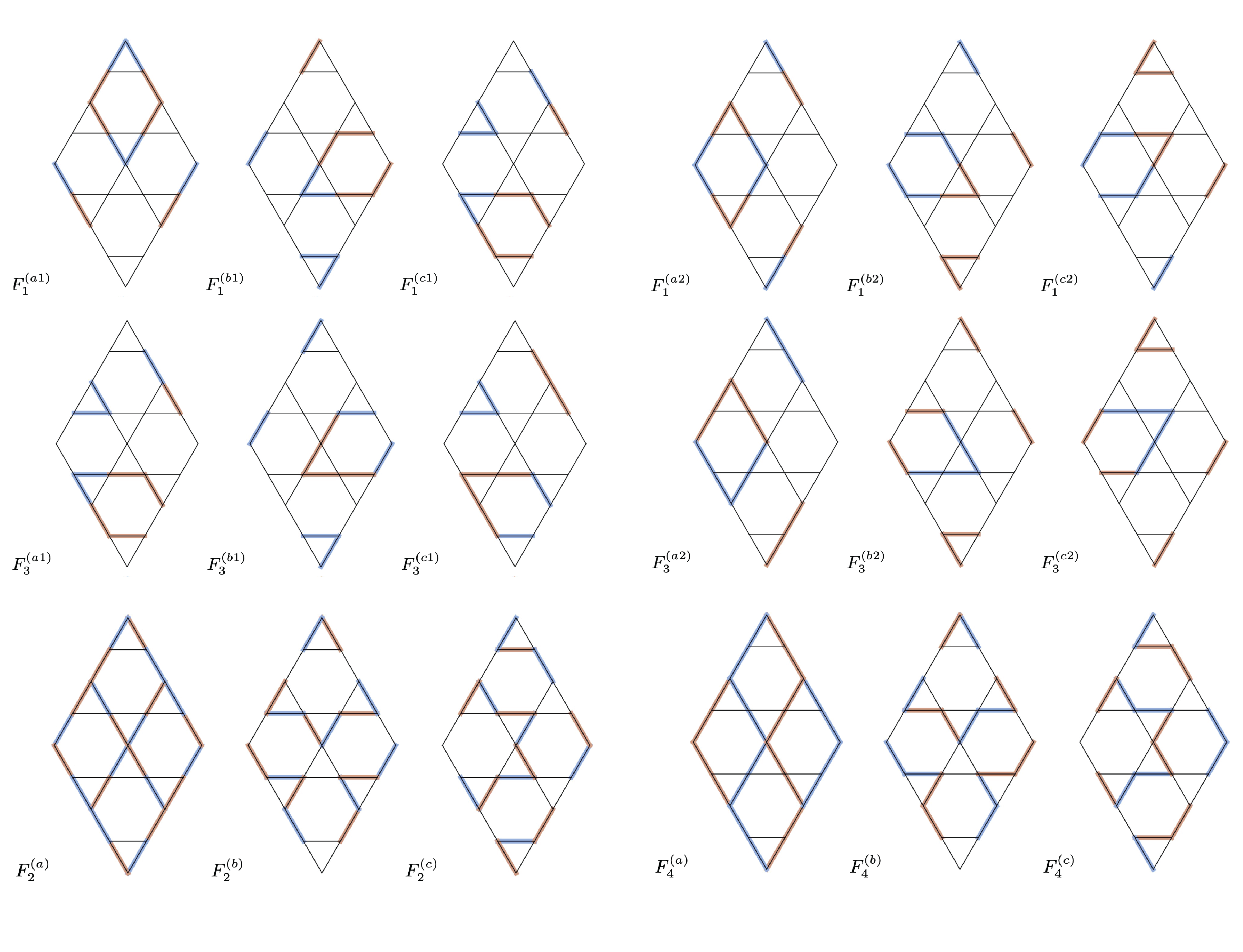}
\vspace{-0.5cm}
\caption{\textbf{Bond orders on the kagome lattice.}  Visual representation of the symmetry-distinct $2\times 2$ bond order functions in the enlarged unit cell. Uncoloured bonds have zero value of the order parameter, blue indicates $+1$ and red $-1$. The notation $F_i^{a,b,c}$ denotes the three distinct components of the $F$ irrep via the upper index.}
\label{f:supp_figX}
\vspace{0cm}
\end{figure*}

Given an order parameter which develops at the Fermi surface, a careful downfolding procedure using the Fermi surface form factors can be used to construct the real space form of the order parameter (see the Supplemental Material of Refs. \cite{li2020artificial, li2021higher, scammell2022intrinsic}). Here for the purposes of illustration, we simply take mean field Hamiltonians using symmetry-allowed basis functions. Specialising to $2\times 2$ bond order CDWs on the kagome lattice, we list a set of basis functions corresponding to each irrep $F_i$. In soon to be posted work, we will apply the same analysis to $\sqrt{3} \times \sqrt{3}$ CDWs.

Bond-ordered charge density waves can be physically interpreted as spatial modulations of the effective nearest-neighbour hopping, and arise naturally as instabilities in kagome Hubbard models \cite{Yu2012, Kiesel2012, Kiesel2013}. The Fourier space basis functions for the four (two) symmetry-distinct bond (flux) orders are given below. The spatial basis functions of each bond order irrep are represented visually in Fig. \ref{f:supp_figX}. \\

\noindent {\bf $\mathbf F_1$ bond order basis}

\[
F_{1}^{\tilde{\mathbf M_1}}(\mathbf k) =\left[F_{1}^{-\tilde{\mathbf M_1}}(\mathbf k)\right]^\dag = \Delta_0 
\begin{pmatrix}
0 & 2 i \sin (\mathbf{k} \cdot \mathbf{d}_{AB}) & -2 i \sin (\mathbf{k} \cdot \mathbf{d}_{CA}) \\
-2 \cos (\mathbf{k} \cdot \mathbf{d}_{AB}) & 0 & 0 \\
-2 \cos (\mathbf{k} \cdot \mathbf{d}_{CA}) & 0 & 0
\end{pmatrix}
\]

\[
F_{1}^{\tilde{\mathbf M_2}}(\mathbf k) =\left[F_{1}^{-\tilde{\mathbf M_2}}(\mathbf k)\right]^\dag = \Delta_0 
\begin{pmatrix}
0 & -2 \cos (\mathbf{k} \cdot \mathbf{d}_{AB}) & 0 \\
-2 i \sin (\mathbf{k} \cdot \mathbf{d}_{AB}) & 0 & 2 i \sin (\mathbf{k} \cdot \mathbf{d}_{BC}) \\
0 & -2 \cos (\mathbf{k} \cdot \mathbf{d}_{BC}) & 0
\end{pmatrix}
\]

\[
F_{1}^{\tilde{\mathbf M_3}}(\mathbf k) =\left[F_{1}^{-\tilde{\mathbf M_3}}(\mathbf k)\right]^\dag = \Delta_0 
\begin{pmatrix}
0 & 0 & -2 \cos (\mathbf{k} \cdot \mathbf{d}_{BC}) \\
0 & 0 & -2 \cos (\mathbf{k} \cdot \mathbf{d}_{CA}) \\
2 i \sin (\mathbf{k} \cdot \mathbf{d}_{CA}) & -2 i \sin (\mathbf{k} \cdot \mathbf{d}_{BC}) & 0
\end{pmatrix}
\]

\noindent {\bf $\mathbf F_2$ bond order basis}

\[
F_{2}^{\tilde{\mathbf M_1}}(\mathbf k) =\left[F_{2}^{-\tilde{\mathbf M_1}}(\mathbf k)\right]^\dag = \Delta_0 
\begin{pmatrix}
0 & 2 i \sin (\mathbf{k} \cdot \mathbf{d}_{AB}) & 2 i \sin (\mathbf{k} \cdot \mathbf{d}_{CA}) \\
2 \cos (\mathbf{k} \cdot \mathbf{d}_{AB}) & 0 & 0 \\
-2 \cos (\mathbf{k} \cdot \mathbf{d}_{CA}) & 0 & 0
\end{pmatrix}
\]

\[
F_{2}^{\tilde{\mathbf M_2}}(\mathbf k) =\left[F_{2}^{-\tilde{\mathbf M_2}}(\mathbf k)\right]^\dag = \Delta_0 
\begin{pmatrix}
0 & -2 \cos (\mathbf{k} \cdot \mathbf{d}_{AB}) & 0 \\
2 i \sin (\mathbf{k} \cdot \mathbf{d}_{AB}) & 0 & 2 i \sin (\mathbf{k} \cdot \mathbf{d}_{BC}) \\
0 & 2 \cos (\mathbf{k} \cdot \mathbf{d}_{BC}) & 0
\end{pmatrix}
\]

\[
F_{2}^{\tilde{\mathbf M_3}}(\mathbf k) =\left[F_{2}^{-\tilde{\mathbf M_3}}(\mathbf k)\right]^\dag = \Delta_0 
\begin{pmatrix}
0 & 0 & 2 \cos (\mathbf{k} \cdot \mathbf{d}_{CA}) \\
0 & 0 & -2 \cos (\mathbf{k} \cdot \mathbf{d}_{BC}) \\
2 i \sin (\mathbf{k} \cdot \mathbf{d}_{CA}) & 2 i \sin (\mathbf{k} \cdot \mathbf{d}_{BC}) & 0
\end{pmatrix}
\]

\noindent {\bf $\mathbf F_3$ bond order basis}

\[
F_{3}^{\tilde{\mathbf M_1}}(\mathbf k) =\left[F_{3}^{-\tilde{\mathbf M_1}}(\mathbf k)\right]^\dag = \Delta_0 
\begin{pmatrix}
0 & 2 i \sin (\mathbf{k} \cdot \mathbf{d}_{AB}) & -2 i \sin (\mathbf{k} \cdot \mathbf{d}_{CA}) \\
2 i \sin (\mathbf{k} \cdot \mathbf{d}_{AB}) & 0 & 0 \\
-2 i \sin (\mathbf{k} \cdot \mathbf{d}_{CA}) & 0 & 0
\end{pmatrix}
\]

\[
F_{3}^{\tilde{\mathbf M_2}}(\mathbf k) =\left[F_{3}^{-\tilde{\mathbf M_2}}(\mathbf k)\right]^\dag = \Delta_0 
\begin{pmatrix}
0 & -2 i \sin (\mathbf{k} \cdot \mathbf{d}_{AB}) & 0 \\
-2 i \sin (\mathbf{k} \cdot \mathbf{d}_{AB}) & 0 & 2 i \sin (\mathbf{k} \cdot \mathbf{d}_{BC}) \\
0 & 2 i \sin (\mathbf{k} \cdot \mathbf{d}_{BC}) & 0
\end{pmatrix}
\]

\[
F_{3}^{\tilde{\mathbf M_3}}(\mathbf k) =\left[F_{3}^{-\tilde{\mathbf M_3}}(\mathbf k)\right]^\dag = \Delta_0 
\begin{pmatrix}
0 & 0 & 2 i \sin (\mathbf{k} \cdot \mathbf{d}_{CA}) \\
0 & 0 & -2 i \sin (\mathbf{k} \cdot \mathbf{d}_{BC}) \\
2 i \sin (\mathbf{k} \cdot \mathbf{d}_{CA}) & -2 i \sin (\mathbf{k} \cdot \mathbf{d}_{BC}) & 0
\end{pmatrix}
\]

\noindent {\bf $\mathbf F_4$ bond order basis}

\[
F_{4}^{\tilde{\mathbf M_1}}(\mathbf k) =\left[F_{4}^{-\tilde{\mathbf M_1}}(\mathbf k)\right]^\dag = \Delta_0 
\begin{pmatrix}
0 & 2 \cos (\mathbf{k} \cdot \mathbf{d}_{AB}) & -2 \cos (\mathbf{k} \cdot \mathbf{d}_{CA}) \\
-2 i \sin (\mathbf{k} \cdot \mathbf{d}_{AB}) & 0 & 0 \\
-2 i \sin (\mathbf{k} \cdot \mathbf{d}_{CA}) & 0 & 0 
\end{pmatrix}
\]

\[
F_{4}^{\tilde{\mathbf M_2}}(\mathbf k) =\left[F_{4}^{-\tilde{\mathbf M_2}}(\mathbf k)\right]^\dag = \Delta_0 
\begin{pmatrix}
0 & -2 i \sin (\mathbf{k} \cdot \mathbf{d}_{AB}) & 0 \\
-2 \cos (\mathbf{k} \cdot \mathbf{d}_{AB}) & 0 & 2 \cos (\mathbf{k} \cdot \mathbf{d}_{BC}) \\
0 & -2 i \sin (\mathbf{k} \cdot \mathbf{d}_{BC}) & 0
\end{pmatrix}
\]

\[
F_{4}^{\tilde{\mathbf M_3}}(\mathbf k) =\left[F_{4}^{-\tilde{\mathbf M_3}}(\mathbf k)\right]^\dag = \Delta_0 
\begin{pmatrix}
0 & 0 & -2 i \sin (\mathbf{k} \cdot \mathbf{d}_{CA}) \\
0 & 0 & -2 i \sin (\mathbf{k} \cdot \mathbf{d}_{BC}) \\
2 \cos (\mathbf{k} \cdot \mathbf{d}_{CA}) & -2 \cos (\mathbf{k} \cdot \mathbf{d}_{BC}) & 0
\end{pmatrix}
\]

\noindent {\bf $\mathbf F_2$ flux order basis}

\[
F_{2}^{\tilde{\mathbf M_1}}(\mathbf k) =F_{2}^{-\tilde{\mathbf M_1}}(\mathbf k) = \Delta_0 
\begin{pmatrix}
0 & 0 & 0 \\
0 & 0 & \sin\left(\mathbf{k} \cdot \mathbf{d}_{BC}\right) \\
0 & \sin\left(\mathbf{k} \cdot \mathbf{d}_{BC}\right) & 0
\end{pmatrix}
\]

\[
F_{2}^{\tilde{\mathbf M_2}}(\mathbf k) =F_{2}^{-\tilde{\mathbf M_2}}(\mathbf k) = \Delta_0 
\begin{pmatrix}
0 & 0 & \sin\left(\mathbf{k} \cdot \mathbf{d}_{CA}\right) \\
0 & 0 & 0 \\
\sin\left(\mathbf{k} \cdot \mathbf{d}_{CA}\right) & 0 & 0
\end{pmatrix}
\]

\[
F_{2}^{\tilde{\mathbf M_3}}(\mathbf k) =F_{2}^{-\tilde{\mathbf M_3}}(\mathbf k) = \Delta_0 
\begin{pmatrix}
0 & \sin\left(\mathbf{k} \cdot \mathbf{d}_{AB}\right) & 0 \\
\sin\left(\mathbf{k} \cdot \mathbf{d}_{AB}\right) & 0 & 0 \\
0 & 0 & 0
\end{pmatrix}
\]

\noindent {\bf $\mathbf F_4$ flux order basis}

\[
F_{4}^{\tilde{\mathbf M_1}}(\mathbf k) =F_{4}^{-\tilde{\mathbf M_1}}(\mathbf k) = i \Delta_0 
\begin{pmatrix}
0 & 0 & 0 \\
0 & 0 & -\cos\left(\mathbf{k} \cdot \mathbf{d}_{BC}\right) \\
0 & \cos\left(\mathbf{k} \cdot \mathbf{d}_{BC}\right) & 0
\end{pmatrix}
\]

\[
F_{4}^{\tilde{\mathbf M_2}}(\mathbf k) =F_{4}^{-\tilde{\mathbf M_2}}(\mathbf k) = i \Delta_0 
\begin{pmatrix}
0 & 0 & \cos\left(\mathbf{k} \cdot \mathbf{d}_{CA}\right) \\
0 & 0 & 0 \\
-\cos\left(\mathbf{k} \cdot \mathbf{d}_{CA}\right) & 0 & 0
\end{pmatrix}
\]

\[
F_{4}^{\tilde{\mathbf M_3}}(\mathbf k) =F_{4}^{-\tilde{\mathbf M_3}}(\mathbf k) = i \Delta_0 
\begin{pmatrix}
0 & -\cos\left(\mathbf{k} \cdot \mathbf{d}_{AB}\right) & 0 \\
\cos\left(\mathbf{k} \cdot \mathbf{d}_{AB}\right) & 0 & 0 \\
0 & 0 & 0
\end{pmatrix}
\]

\subsubsection{Translationally invariant orders}

In addition, we define a translationally invariant $B_1$ bond order, whose presence should be detectable using the Bragg peaks at the reciprocal lattice vectors. \\

\noindent {\bf ${\mathbf B_1}$ bond order basis}

\[
B_1(\mathbf{k}) = -i \Delta_0 \begin{pmatrix}
0 & \sin\left(\mathbf{k} \cdot \mathbf{d}_{AB}\right) & -\sin\left(\mathbf{k} \cdot \mathbf{d}_{CA}\right) \\
-\sin\left(\mathbf{k} \cdot \mathbf{d}_{AB}\right) & 0 & \sin\left(\mathbf{k} \cdot \mathbf{d}_{BC}\right) \\
\sin\left(\mathbf{k} \cdot \mathbf{d}_{CA}\right) & -\sin\left(\mathbf{k} \cdot \mathbf{d}_{BC}\right) & 0
\end{pmatrix}
\]

\subsection{Results: synthetic STM data with and without CDW order}

\begin{figure*}[t]
\centering
\includegraphics[width = \textwidth]{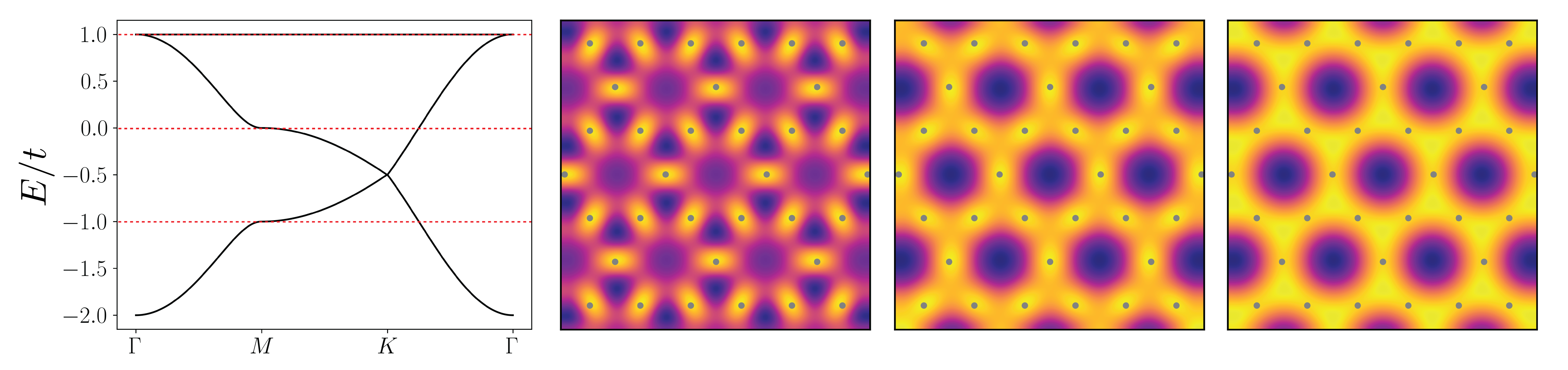}	
\caption{\textbf{Charge density in the kagome tight-binding model.} (a) Bandstructure of the nearest-neighbor kagome tight-binding model, featuring a flat band at $E/t=1$, a $p$-type Van Hove singularity at $E/t = 0$, and an $m$-type Van Hove singularity at $E/t=-1$, each highlied by red dashed lines. The flat band wave function is localised in the center of the hexagons, the $p$-type wavefunction is localised on a different distinct sublattice for each $M$-point, and the $m$-type wavefunction is localised on a different pair of sublattices for each $M$-point. (b) $\text{LDOS}(E=t, \mathbf{r})$; the flat band eigenstate contributes DOS at the center of the hexagons, while the elongated DOS on the sites derive from the dispersive band which touches the flat band at $\Gamma$. (b) $\text{LDOS}(E=0, \mathbf{r})$; unsurprisingly, the $p$-type LDOS is maximal on the atomic sites. (c) $\text{LDOS}(E=-t, \mathbf{r})$; due to the destructive interference between the $m$-type Bloch states on neighbouring sites, the maxima of the charge density form a honeycomb lattice.}
\label{f:tb_ldos}
\vspace{-0.2cm}
\end{figure*}

We begin by presenting results for the LDOS in the absence of charge order, i.e. computed simply from the nearest-neighbour tight-binding model; the results are shown in Fig. \ref{f:tb_ldos}. The positions of maximal charge density are determined by the interference between neighbouring orbitals convolved with their Bloch wavefunctions at a given momentum \cite{holbrook2024real} -- the flatband wavefunction contributes density of states near the center of the hexagon, and is degenerate with a dispersive band whose contribution to the density of states lies on bonds between the triangular network of flatband states. The $m$- and $p$-type charge density form a kagome and honeycomb lattice respectively.

\begin{figure*}[t]
\centering
\includegraphics[width = \textwidth]{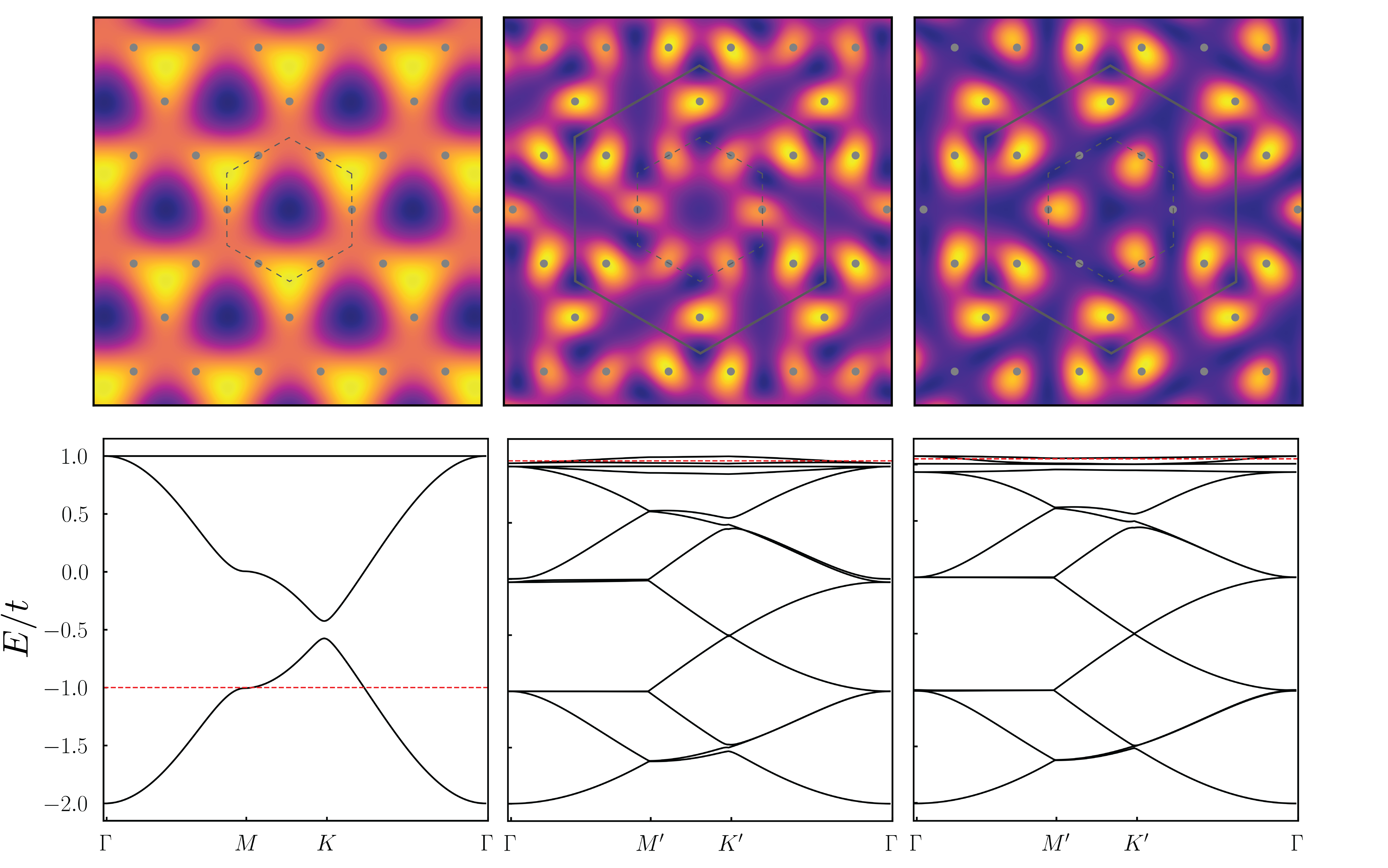}	
\caption{{\bf Local density of states of the charge-ordered kagome lattice.} Bandstructure (bottom) and $\text{LDOS}(E,\mathbf{r})$ (top) of the kagome model with various order parameters: (a) $F_4$ flux order with $\Delta_0/t=0.1$, and $E/t=2.1$. (b)  $F_2$ bond order with $\Delta_0/t=0.1$ and $E/t=2.1$.  (c)  $B_1$ bond order with $\Delta_0/t=0.2$ and $E/t=-2.0$. Solid (dashed) hexagon shows the $2\times2$ CDW unit cell (original unit cell). }
    \label{fig:STM_synth}
\vspace{-0.2cm}
\end{figure*}

Perturbing the tight-binding model with various order parameters, in Fig. \ref{fig:STM_synth} we plot the resulting bandstructure for $B_1$ and $F_2$ bond orders, and $F_4$ flux order, along with the LDOS at different energies indicated by the red dashed lines. The breaking of $C_2$ and mirror symmetries in the appropriate cases are clearly observable, while the pattern of charge density can be understood as a symmetry breaking deformation of the unperturbed charge density -- for instance, for the $B_1$ order near the $m$-type Van Hove singularity, we find a honeycomb lattice with one sublattice more densely populated than the other.

\subsection{Results: symmetry decomposition of synthetic data}\label{sec:window_synth}

In this section, we apply the various symmetry decomposition techniques to synthetic STM data. These include Bragg peak decompositions at both the first and third $M$-point shells, as well as the real-space decomposition approach. Each scheme is also tested in conjunction with the symmetry convolution method, allowing us to assess its utility in imaging irreps that are extinct in conventional analysis.

\subsubsection{Bragg peak decomposition}

We begin by benchmarking our method using synthetic data constructed from an $F_1$ bond-order. The parameters used are: Length = 8, GridPoints = 410881, and $\Delta=0.1t$. In Table~\ref{tab:F1_M3_Bragg}, we present the symmetry decomposition results using the third shell of Bragg peaks, as a function of the window function spread $\sigma$. For large $\sigma$ in the window function Eq. \eqref{window_func}, i.e., weak windowing, we observe spurious contributions to the $F_2$ channel as well as artificial $C_3$ symmetry breaking in the $F_1$ components. These artefacts are systematically suppressed as $\sigma$ is reduced, with optimal suppression observed around $\sigma=2.0$. Note: here the value of $\sigma$ is measured relative to the total number of points along the $x$-axis, call it $n_x$, i.e. choosing $N$ in $\sigma = (n_x-1)/N$.

\begin{table*}[b!]
\begin{center}
\caption{\textbf{Bragg peak decomposition of $F_1$ data as a function of window function spread $\sigma$.} The vectors $\Phi^{(3)}_{F_i}$ represent the symmetry-decomposed STM data from the third shell of Bragg peaks, Eqs. \eqref{F1_3}-\eqref{F4_3}. }
\label{tab:F1_M3_Bragg}
\vspace{0.1cm}
\begin{ruledtabular}
  \begin{tabular}{ccccc} \\[-2.0mm]
     $\sigma$  & $\Phi^{(3)}_{F_1}$           & $\Phi^{(3)}_{F_2}$            & $\Phi^{(3)}_{F_3}$            & $\Phi^{(3)}_{F_4}$            \\[2mm] 
    \hline \vspace{-0.0cm} \\ \vspace{0.2cm}
    100.0  & $\{-0.521,\,-0.604,\,-0.604\}$ & $\{0.000,\,-0.100,\,0.099\}$   & $\{0.000,\,0.000,\,0.000\}$    & $\{0.000,\,0.000,\,0.000\}$    \\   \vspace{0.2cm}
    4.0    & $\{-0.528,\,-0.601,\,-0.601\}$ & $\{0.000,\,-0.045,\,0.044\}$   & $\{0.000,\,0.000,\,0.000\}$    & $\{0.000,\,0.000,\,0.000\}$    \\   \vspace{0.2cm}
    2.0    & $\{-0.558,\,-0.588,\,-0.588\}$ & $\{0.000,\,-0.011,\,0.010\}$   & $\{0.000,\,0.000,\,0.000\}$    & $\{0.000,\,0.000,\,0.000\}$    \\[1mm]
  \end{tabular}
\end{ruledtabular}
\end{center}
\end{table*}

We next analyse synthetic data corresponding to non-trivial order parameter irreps. The parameters are held fixed from the $F_1$ case: GridPoints = 410881 and $\Delta=0.1t$. The $F_2$ and $F_3$ cases correspond to bond orders, while $F_4$ is implemented as a flux order. Table~\ref{tab:decomp_Fi} shows the results of Bragg peak decomposition using the third Brillouin zone, with a window function of $\sigma=2.0$. As expected from theoretical considerations, each $F_i$ order induces a signal in the $F_1$ channel as well as its own irrep; the $F_1$ component reflects the trivial symmetry preserved in the LDOS, while the presence of $F_i$ confirms the correct irrep content.

\begin{table}[t!]
\begin{center}
\renewcommand{\arraystretch}{1.5}
\caption{\textbf{Bragg peak decomposition in the third Brillouin zone for distinct irreps.} A window function of $\sigma = 2.0$  is used throughout. One finds that an order parameter $F_i$  produces symmetry signatures in the LDOS in the $F_1$ and $F_i$ channels. }
\label{tab:decomp_Fi}
\vspace{0.1cm}
\begin{ruledtabular}
  \begin{tabular}{ccccc} \\[-2.0mm]
    \textbf{CDW} & $\Phi^{(3)}_{F_1}$           & $\Phi^{(3)}_{F_2}$            & $\Phi^{(3)}_{F_3}$            & $\Phi^{(3)}_{F_4}$            \\[2mm]
    \hline \vspace{-0.0cm} \\ \vspace{0.2cm}
    $F_2$ & $\{0.125,\;0.121,\;0.142\}$    & $\{0.557,\,-0.584,\,-0.589\}$ & $\{0.000,\;0.000,\;0.000\}$   & $\{0.000,\;0.000,\;0.000\}$   \\   \vspace{0.2cm}
    $F_3$ & $\{-0.556,\,-0.587,\,-0.587\}$ & $\{0.000,\,-0.010,\;0.010\}$  & $\{-0.018,\;0.019,\;0.019\}$   & $\{0.000,\;0.000,\;0.000\}$   \\   \vspace{0.2cm}
    $F_4$ & $\{0.171,\;0.180,\;0.180\}$    & $\{0.000,\;0.003,\,-0.003\}$  & $\{0.000,\,-0.010,\;0.010\}$  & $\{-0.557,\,-0.587,\,-0.587\}$ \\[1mm]
  \end{tabular}
\end{ruledtabular}
\end{center}
\end{table}

We now demonstrate the utility of the symmetry convolution method, using Bragg peaks in the first Brillouin zone. In Table~\ref{tab:bragg_mask}, we convolve the synthetic STM data with symmetry masks $B_i$ before decomposition. As anticipated, the convolution enhances the signal of specific irreps that are otherwise extinct in the first BZ. This confirms that the symmetry mask technique effectively enables access to symmetry channels normally absent due to extinction constraints.

\begin{table}[t!]
\begin{center}
\renewcommand{\arraystretch}{1.5}
\caption{\textbf{Bragg peak decomposition with symmetry masks.} Bragg-peak decomposition method for different order parameters, with LDOS treated via the symmetry convolution method. A window function with $\sigma = 2.0$ is used in each case.}
\label{tab:bragg_mask}
\vspace{0.1cm}
\begin{ruledtabular}
  \begin{tabular}{cccc} \\[-2.0mm]
    CDW & Symmetry mask & $\Phi^{(1)}_{F_1}$                       & $\Phi^{(1)}_{F_3}$                       \\[2mm]
    \hline \vspace{-0.0cm} \\[-2mm]
    $F_2$ & None   & $\{-0.585,\,-0.572,\,-0.573\}$ & $\{0.000,\;0.000,\;0.000\}$ \\   \vspace{0.2cm}
          & $B_2$  & $\{0.000,\;0.000,\;0.000\}$     & $\{-0.715,\,-0.700,\,-0.700\}$ \\[1mm]
    \hline \vspace{-0.0cm} \\[-2mm]
    $F_3$ & None   & $\{-0.584,\,-0.573,\,-0.573\}$ & $\{-0.493,\,-0.482,\,-0.482\}$ \\   
          & $B_1$  & $\{-0.142,\,-0.138,\,-0.138\}$ & $\{3.114,\,3.043,\,3.043\}$     \\  \vspace{0.2cm}
          & $B_2$  & $\{0.000,\,0.000,\,0.000\}$     & $\{0.000,\,0.000,\,0.000\}$     \\[1mm]
    \hline \vspace{-0.0cm} \\[-2mm]
    $F_4$ & None   & $\{-0.586,\,-0.572,\,-0.572\}$ & $\{0.000,\;0.000,\;0.000\}$ \\   \vspace{0.2cm}
          & $B_2$  & $\{0.828,\;0.808,\;0.808\}$     & $\{0.000,\;0.000,\;0.000\}$ \\[1mm]
  \end{tabular}
\end{ruledtabular}
\end{center}
\end{table}

\subsubsection{Real space decomposition}

We compare the output of the Bragg peak decomposition with the results we find from the real space method. Beginning with $F_4$ fluc order, the latter yields (normalised so that $|(\bar{{\cal W}}_{F_{1a}},\bar{{\cal W}}_{F_{1b}})|=1$)
\begin{align}
\begin{pmatrix} \mathcal{W}_{\mathbf{M}_1} \\ \mathcal{W}_{\mathbf{M}_2} \\ \mathcal{W}_{\mathbf{M}_3} \end{pmatrix}_{F_{1a}} & = \begin{pmatrix}
    0.0411368\\ 0.0411292\\ 0.0411429
\end{pmatrix}, \quad
\begin{pmatrix} \mathcal{W}_{\mathbf{M}_1} \\ \mathcal{W}_{\mathbf{M}_2} \\ \mathcal{W}_{\mathbf{M}_3} \end{pmatrix}_{F_{1b}}  = \begin{pmatrix}
    0.575882 \\ 0.575892 \\ 0.575875
\end{pmatrix}, \quad 
\begin{pmatrix} \mathcal{W}_{\mathbf{M}_1} \\ \mathcal{W}_{\mathbf{M}_2} \\ \mathcal{W}_{\mathbf{M}_3} \end{pmatrix}_{F_{4}} = \begin{pmatrix}
    1.03927 \\ 1.03927 \\ 1.03928
\end{pmatrix}
\end{align}
and $|\bar{{\cal W}}_{F_{2}}|\approx |\bar{{\cal W}}_{F_{3a}}| \approx|\bar{{\cal W}}_{F_{3b}}|=0 + O(10^{-5})$. Meanwhile, for $F_2$ bond order
\begin{align}
\begin{pmatrix} \mathcal{W}_{\mathbf{M}_1} \\ \mathcal{W}_{\mathbf{M}_2} \\ \mathcal{W}_{\mathbf{M}_3} \end{pmatrix}_{F_{1a}} &= \begin{pmatrix}
    -0.463051 \\ -0.463046 \\ -0.463046
\end{pmatrix}, \quad
\begin{pmatrix} \mathcal{W}_{\mathbf{M}_1} \\ \mathcal{W}_{\mathbf{M}_2} \\ \mathcal{W}_{\mathbf{M}_3} \end{pmatrix}_{F_{1b}}  = \begin{pmatrix}
    0.344853 \\ 0.344846 \\ 0.344846
\end{pmatrix}, \quad
\begin{pmatrix} \mathcal{W}_{\mathbf{M}_1} \\ \mathcal{W}_{\mathbf{M}_2} \\ \mathcal{W}_{\mathbf{M}_3} \end{pmatrix}_{F_{2}} = \begin{pmatrix}
    3.30148 \\ 3.30148 \\ 3.30148
\end{pmatrix},
\end{align}
and $|\bar{{\cal W}}_{F_{3a}}|\approx |\bar{{\cal W}}_{F_{3b}}| \approx|\bar{{\cal W}}_{F_{4}}|=0 + O(10^{-5})$. That is, both methods conform perfectly with our theoretical predictions.

\begin{figure}[b]
    \centering
\includegraphics[width=0.58\textwidth]{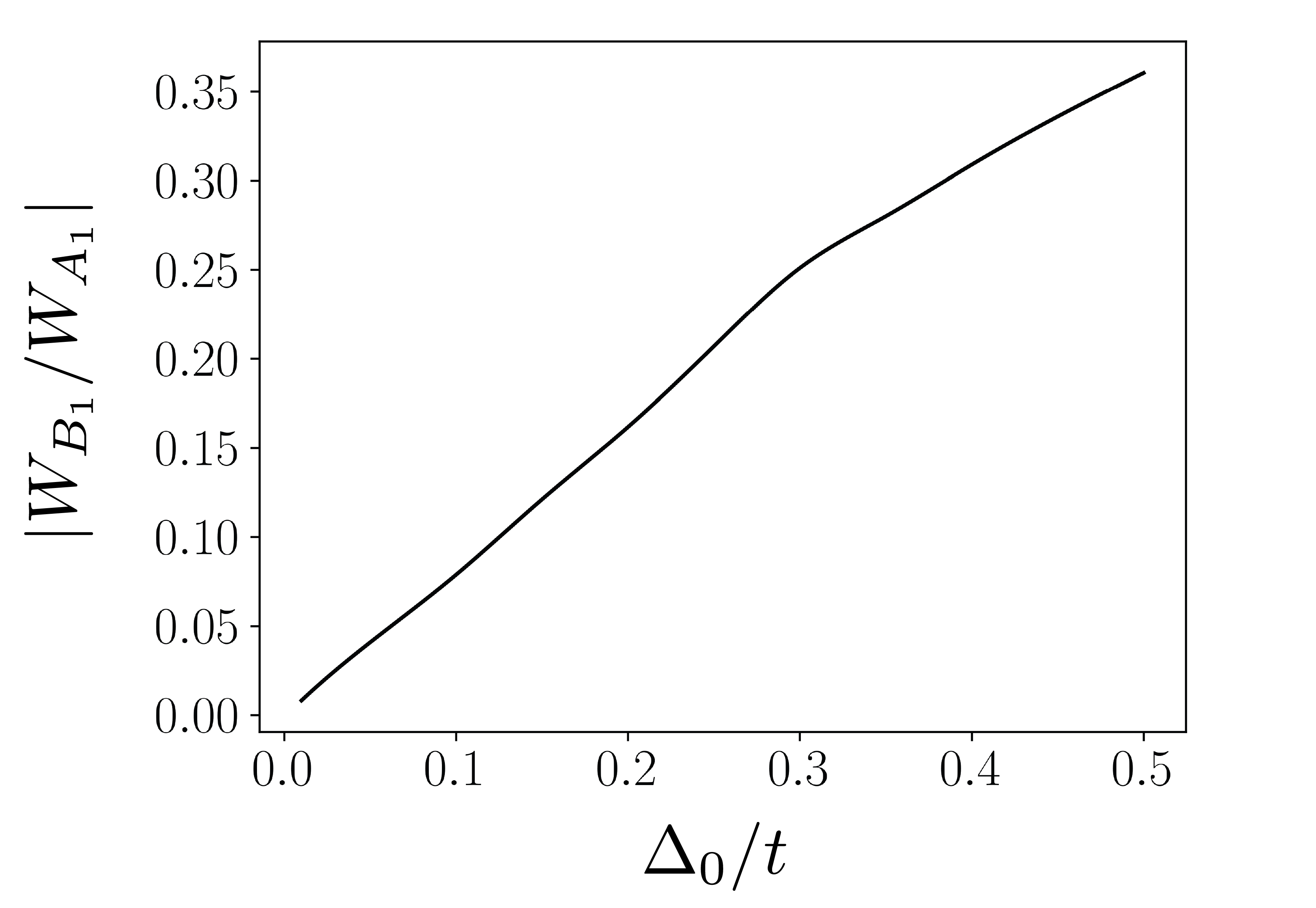}
    \vspace{-0.2cm}
    \caption{\textbf{LDOS signal as a function of $B_1$ order parameter strength.} Ratio of the weight of $B_1$ to $A_1$ irreps versus the magnitude of the BdG order parameter, $\Delta_0/t$, obtained from decomposition of $\text{LDOS}(E=-2,\mathbf{r})$ for translationally invariant order $B_1$.}
    \label{fig:WB1_Delta}
\end{figure}

Lastly, we analyse translationally invariant order. Here we take $B_1$ bond order as our example; synthetic STM is shown in Fig. \ref{fig:STM_synth}(c), and the real space decomposition yields that ${\cal W}_{B_1}/{\cal W}_{A_1}\propto \Delta_0/t$ for  $\Delta_0/t\lesssim0.25$; see Fig. \ref{fig:WB1_Delta}.

\clearpage

\section{Demonstration on ScV$_6$Sn$_6$ topography}\label{supp-scv6sn6}

\subsection{Objective}
In this section, we turn to real STM topographic data. Our objective is to demonstrate the data handling required to prepare the data for Fourier decomposition analysis. We do not attempt to establish properties of the ScV$_6$Sn$_6$ phase diagram. ScV$_6$Sn$_6$ has hexagonal symmetry and is known to form a high-temperature $\sqrt{3}\times\sqrt{3}$ CDW. We therefore decompose the STM signal according to the irreps of $C_{6v}^{\prime\prime}$, detailed in Table~\ref{tab:cpp6v}. Since we do not have control over which irreps appear in the data, we run the decomposition on two forms of the STM signal: (i) the original data, subject to appropriate preprocessing such as a shift of origin and rescaling for clarity; and (ii) a symmetrised version of the data, constructed by averaging over all six-fold rotations to impose approximate $C_6$ symmetry. The windowed real-space data, with and without $C_6$-symmetrisation, and the subsequent FFT are shown in Fig.~\ref{fig:FFTSTM_manip}.

\subsection{Analysis}\label{sec:window}
The results of the STM symmetry decomposition are summarised in the first row of Table~\ref{tab:Eg_irrep_decomp}. Two cases are shown: (i) the raw STM data with a shifted origin, and (ii) the same data with approximate $C_6$ symmetry imposed. In both cases, the extracted order parameter components in the $E'_1$ and $G$ irreps are reported, as well as the Gaussian width of the window function, $\sigma$. The norm of each irrep component is also listed for quantitative comparison.

If the data is truly $C_6$-symmetric, then for the $\sqrt{3}\times\sqrt{3}$ CDW irreps, one expects
\begin{align}
\text{$C_6$-symmetry:} \quad \mathbf \Phi_{E_1^{'}}=\{1,0\}, \quad \mathbf \Phi_{G}=\{0,0,0,0\},
\end{align}
which follows from Eqs.~\eqref{Ep1} and \eqref{G}. Note here we take normalisation $|\mathbf \Phi_{E_1^{'}}|=1$. Similarly, for translationally invariant irreps, only $\Phi_{A_1}\neq0$, while $\Phi_{B_1}=|\mathbf \Phi_{E_1}|=|\mathbf\Phi_{E_2}|=0$; see Eqs.~\eqref{A1}, \eqref{B1}, \eqref{E1} and \eqref{E2}. We evaluate the translationally invariant irreps as shown in the first row of Table~\ref{tab:AB_irrep_decomp}.

For the $C_6$-symmetrised data, we find $|\Phi_{G}|/|\Phi_{E'_1}| = 0.02$ and $|\Phi_{E_2}|/|\Phi_{A_1}| = 1.6\times 10^{-15}$, indicating that the windowing has sufficiently removed spurious rotational symmetry breaking.

\begin{table}[h]
\begin{center}
\renewcommand{\arraystretch}{1.5}
\setlength{\tabcolsep}{10pt}
\caption{\textbf{Decomposition into $E'_1$ and $G$ irreps under both raw and $C_6$‐symmetric conditions.} The final column shows the magnitude ratio $|\Phi_G|/|\Phi_{E'_1}|$. The data is on $(256,256)$ grid and $\sigma=21$ pixels.}
\label{tab:Eg_irrep_decomp}
\vspace{0.1cm}
\begin{ruledtabular}
  \begin{tabular}{lccc} \\[-2.0mm]
     \textbf{Condition} & $\boldsymbol{\Phi_{E'_1}}$      & $\boldsymbol{\Phi_G}$                         & $\boldsymbol{|\Phi_G|/|\Phi_{E'_1}|}$ \\[2mm]
    \hline \\[-2mm]
     $C_6$              & $\{-1.000,\,-0.000\}$           & $\{ -0.011,\ 0.008,\ -0.011,\ -0.009 \}$         & 0.019                                \\[0.2cm]
     Raw                & $\{ -0.534,\ 0.845 \}$           & $\{ 0.239,\ 0.078,\ 0.170,\ 0.066 \}$          & 0.311                                     \\[1mm]
  \end{tabular}
\end{ruledtabular}
\end{center}
\end{table}
\begin{table}[h]
\begin{center}
\renewcommand{\arraystretch}{1.4}
\setlength{\tabcolsep}{8pt}
\caption{\textbf{Fourier decomposition into $C_{6v}$ irreducible representations using G vectors.} The $A_1$ irrep remains dominant, while $|\Phi_{E_2}|/|\Phi_{A_1}|$ (last column) quantifies symmetry‐breaking components. The data is on $(256,256)$ grid and $\sigma=21$ pixels.}
\label{tab:AB_irrep_decomp}
\vspace{0.1cm}
\begin{ruledtabular}
  \begin{tabular}{lccccc} \\[-2.0mm]
     \textbf{Condition} & $\boldsymbol{\Phi_{A_1}}$    & $\boldsymbol{\Phi_{B_1}}$    & $\boldsymbol{\Phi_{E_1}}$          & $\boldsymbol{\Phi_{E_2}}$          & $\boldsymbol{|\Phi_{E_2}|/|\Phi_{A_1}|}$ \\[2mm]
    \hline \\[-2mm]
    $C_6$              & $-1.000$                     & $0.000$                      & $\{ -0.017,\ 0.010\}$            & $\{ 0.000,\ -0.001 \}$                & $1.56\times 10^{-15}$                                 \\[0.2cm]
   Raw                & $-1.000$                     & $-0.007$                      & $\{ -0.016,\ 0.000 \}$             & $\{ 0.100,\ -0.066 \}$                & 0.120                                 
\\[2mm]
  \end{tabular}
\end{ruledtabular}
\end{center}
\end{table}

\begin{figure}[t!]
\centering
\includegraphics[width=0.85\textwidth]{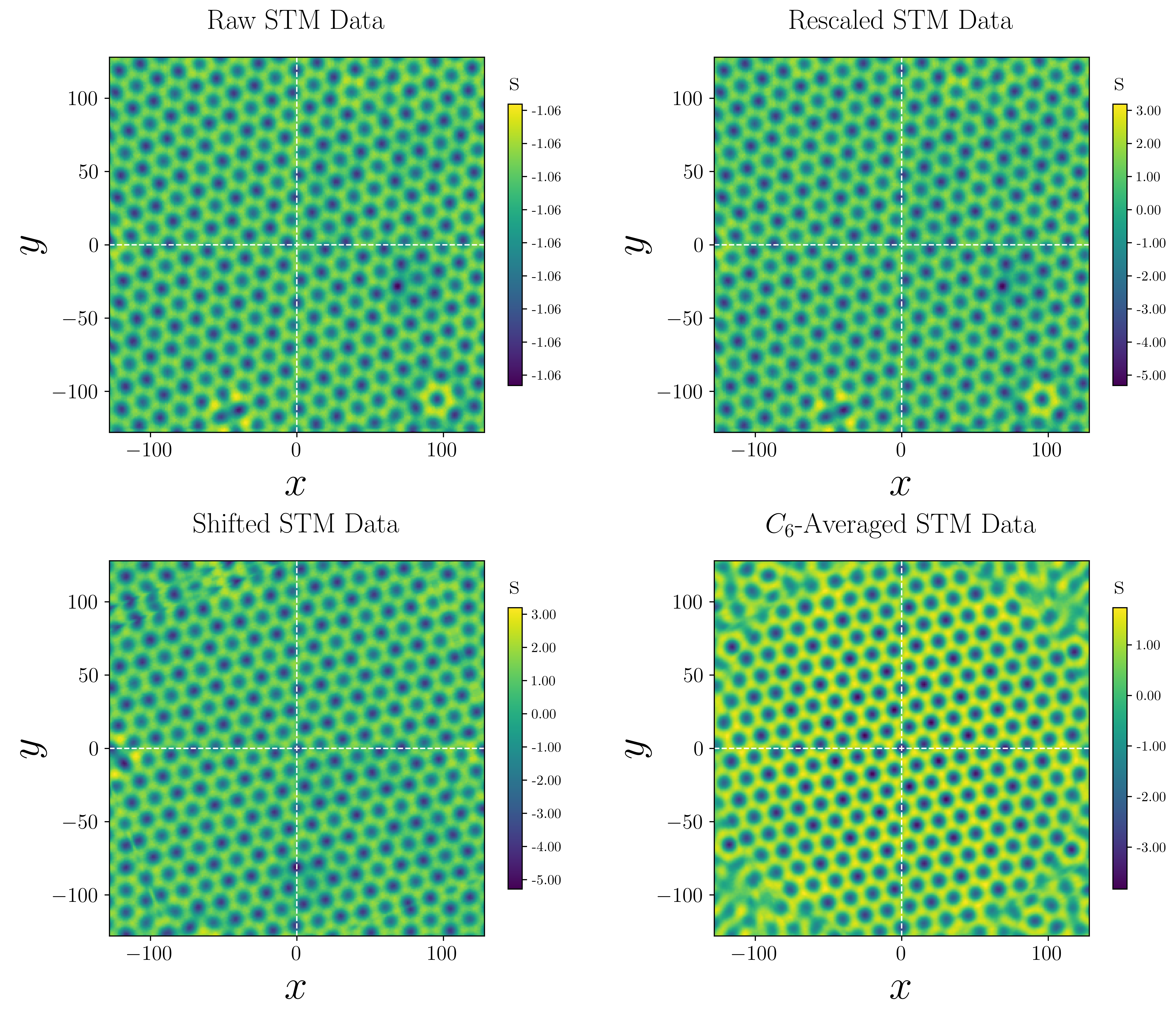}
\caption{\textbf{Manipulating real topographic data for ScV$_6$Sn$_6$.} $C_6$-rotations of the data creates artificial domain-like walls in the corners of the $(x,y)$ plane. Employing a window function removes those spurious features.}
    \label{fig:realSTM_manip}
\end{figure}

\begin{figure}[t!]
\centering
\includegraphics[width=0.85\textwidth]{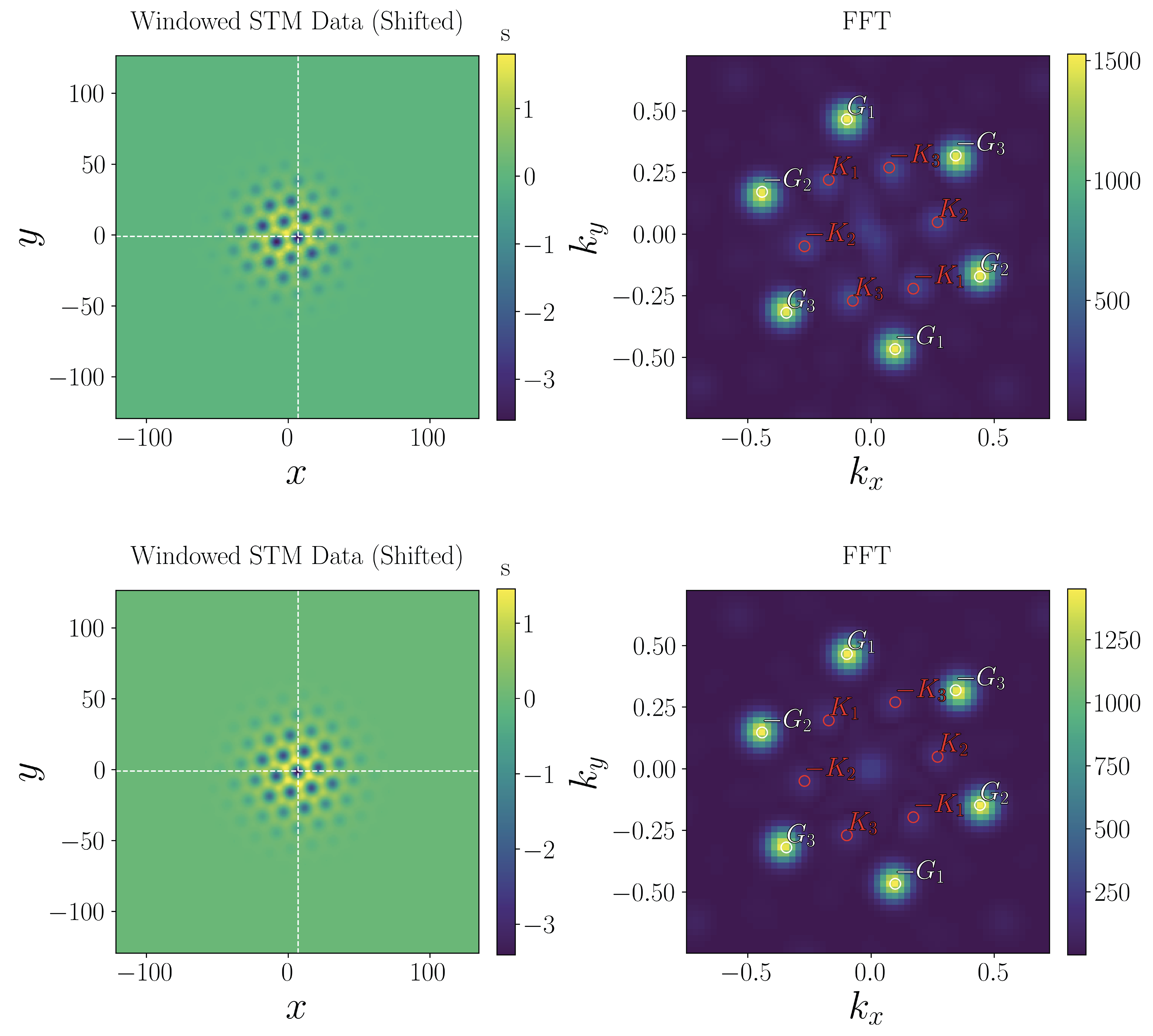}
\caption{\textbf{Windowed STM and FFT.} (Top) STM data, with shifted centre and rescaling. (Bottom) $C_6$-symmetry imposed on real STM data, also with shifted centre and rescaling.}
    \label{fig:FFTSTM_manip}
\end{figure}

\end{document}